\documentclass[fleqn,usenatbib]{mnras}

\usepackage[T1]{fontenc}
\usepackage{colortbl}
\usepackage{tabularx}
\usepackage{placeins}
\usepackage{float}
\usepackage{lipsum}
\usepackage{xcolor}
\usepackage{caption}
\usepackage{subcaption}
\usepackage{soul}
\usepackage[export]{ adjustbox }

\DeclareRobustCommand{\VAN}[3]{#2}
\let\VANthebibliography\thebibliography
\def\thebibliography{\DeclareRobustCommand{\VAN}[3]{##3}\VANthebibliography}

\usepackage{graphicx}	
\usepackage{amsmath}	
\usepackage{comment}
\usepackage{babel}
\usepackage{newtxtext,newtxmath}

\title[Galactic Bulge Radio Survey]{An arcsecond view at 1--2~GHz into the Galactic Bulge}

\author[E. C. Pattie et al.]{ 
E. C. Pattie,$^{1}$\thanks{E-mail: eli.pattie@ttu.edu} 
T. J. Maccarone,$^{1}$
C. T. Britt,$^{2}$ 
C. O. Heinke,$^{3}$ 
P. G. Jonker,$^{4,5}$ 
D. R. Lorimer,$^{6,7}$ 
\newauthor G. R. Sivakoff,$^{3}$ 
D. Steeghs,$^{8}$ 
J. Strader,$^{9}$ 
M. A. P. Torres$^{10,11}$ 
and R. Wijnands$^{12}$  \\
\\
$^{1}$Department of Physics and Astronomy, Texas Tech University, Lubbock TX 79409, USA \\
$^{2}$Space Telescope Science Institute, 3700 San Martin Drive, Baltimore, Maryland 21218, USA \\
$^{3}$Department of Physics, University of Alberta, Edmonton, AB T6G 2E1, Canada \\
$^{4}$Department of Astrophysics/IMAPP, Radboud University Nĳmegen, P.O. Box 9010, 6500 GL, Nĳmegen, The Netherlands \\
$^{5}$SRON, Netherlands Institute for Space Research, Niels Bohrweg 4, 2333 CA, Leiden, The Netherlands \\
$^{6}$West Virginia University, Department of Physics and Astronomy, P. O. Box 6315,
Morgantown, WV, USA \\
$^{7}$Centre for Gravitational Waves and Cosmology, West Virginia University, Chestnut Ridge Research Building, Morgantown, WV, USA \\
$^{8}$Department of Physics, University of Warwick, Coventry CV4 7AL, UK \\
$^{9}$Centre for Data Intensive and Time Domain Astronomy, Department of Physics and Astronomy, Michigan State University, East Lansing, MI 48824, USA \\
$^{10}$Instituto de Astrofísica de Canarias (IAC), Vía Láctea s/n, La Laguna 38205, S/C de Tenerife, Spain \\
$^{11}$Departamento de Astrofísica, Universidad de La Laguna, La Laguna, E-38205, S/C de Tenerife, Spain \\
$^{12}$Anton Pannekoek Institute for Astronomy, University of Amsterdam, Postbus 94249, 1090 GE Amsterdam, The Netherlands \\
}
\date{Accepted XXX. Received YYY; in original form ZZZ}

\pubyear{2023}

\begin{document}
\label{firstpage}
\pagerange{\pageref{firstpage}--\pageref{lastpage}}
\maketitle

\begin{abstract}
We present the results of a high angular resolution (1.1") and sensitivity (maximum of $\sim$0.1~mJy) radio survey at 1--2~GHz in the Galactic Bulge. This complements the X-ray {\it Chandra} Galactic Bulge Survey, and investigates the full radio source population in this dense Galactic region. Radio counterparts to sources at other wavelengths can aid in classification, as there are relatively few types of objects that are reasonably detectable in radio at kiloparsec distances, and even fewer that are detected in both X-rays and radio. This survey covers about 3 square degrees of the Galactic Bulge Survey area (spanning the Galactic coordinate range of $-3^{\circ}<l<+3^{\circ}$ and $+1.6^{\circ}<b<+2.1^{\circ}$) as a first look into this region of the Galaxy with this combination of frequency, resolution, and sensitivity. Spectral indices within the observed band of 1--2~GHz were calculated for each source to assist in determining its emission mechanism. We find 1617 unique sources in the survey, 25 of which are radio counterparts to X-ray sources, and about 100 of which are steep-spectrum ($\alpha \lesssim -1.4$) point sources that are viable pulsar candidates. Four radio sources are of particular interest: a compact binary; an infrared transient with an inverted radio spectrum; a potential transitional millisecond pulsar candidate; and a very steep spectrum radio source with an X-ray and bright infrared counterpart. We discuss other notable sources, including possible radio transients, potential new planetary nebulae, and active galactic nuclei.
\end{abstract}




\section{Introduction}

All-sky radio surveys offer a glimpse into the radio sky, viewing both Galactic and extragalactic sources. Recent surveys (e.g. Rapid ASKAP Continuum Survey (RACS) \citep{McConnell2020} and Very Large Array Sky Survey (VLASS) \citep{Lacy2020}, see Table~\ref{table:Catalogs}) have been performed with higher angular resolution and sensitivity than their predecessors decades before 
(e.g., the angular resolution/RMS of 2"/0.12~mJy of VLASS and 25"/0.25~mJy of RACS compared to the 1990s NRAO VLA Sky Survey (NVSS) \citep{Condon1998} at 45"/0.75~mJy), revealing many more radio sources overall, and often offering a more detailed view of their morphologies.
Other radio survey projects and large efforts to date have focused on specific targets, e.g., globular clusters \citep{Shishkovsky2020} or molecular clouds \citep{2022ApJ...925...39T}. Though globular clusters can host a large number of sources, they evolve differently compared to the Galaxy and thus host different fractions of objects \citep{Pooley2003}, and therefore are not representative of the Milky Way in these ways. 
Molecular cloud surveys generally focus on young stellar objects. 
Other efforts target sources that were first found at other wavelengths, e.g. masers or star-forming regions \citep{Urquhart2011}. The densely populated regions toward the Galactic centre overall lack comprehensive deep radio imaging efforts, with large surveys to date tending toward low angular resolutions to detect large-scale structures in the interstellar medium \citep[e.g.,][]{Helfand2006MAGPIS,Heywood2022}, rather than resolving point source emission that originates from individual stellar objects. 

There are many Galactic object classes that are particularly important for understanding stellar evolution, such as young stellar objects, planetary nebulae, stellar remnants, and interacting and compact binaries. Many of these can produce observable levels of radio emission at kiloparsec-scale distances in reasonable integration times, which may aid in detecting and identifying them, as there are relatively few types of sources overall that produce such levels of radio emission (due to, e.g., nonthermal processes or large angular sizes). 
These Galactic objects may also be difficult to identify at other wavelengths, due to their faintness or because of crowding issues, which can be especially problematic in the infrared. Thus, radio observations can be quite useful in detecting and classifying certain types of sources in the Galaxy.

\subsection{Pulsars}

Pulsars are a subset of neutron stars that are observed to produce radio emission near their magnetic poles, and their magnetic axes are generally misaligned with respect to their rotation axes. This misalignment results in the appearance of pulsed emission when the magnetic axis sweeps over the Earth's line of sight \citep{Gold1968} as the neutron star spins at frequencies typically on orders of $\sim$milliseconds to seconds \citep{Sturrock1971}. 
Pulsars with spin periods of $\sim$10ms or less are classified as millisecond pulsars (MSPs) \citep{Backer1982}, which are spun up to these very short spin periods by accreting material from a low mass companion in a binary \citep{Alpar1982,Bhattacharya1991}. Most pulsars are steep spectrum sources \citep{Bates2013}, meaning their flux density increases with decreasing frequency (until a turnover frequency, most often observed to occur at $\sim$100~MHz \citealt{Sieber1973,Izvekova1981}). Their radio emission is modelled with a power law: their flux density $S_{\nu} \propto \nu^{\alpha}$, where $\nu$ is the frequency and $\alpha$ is the spectral index. The only method that confirms a radio source as a pulsar is to detect their pulsations with timing observations.

Detecting pulsars through timing observations encounter two significant issues of dispersion (frequency-dependent time delay of the signal due to the interstellar medium $\propto \nu^{-2}$) and scattering (pulse profile broadening due to interstellar material interaction $\propto \nu^{-4}$) \citep{Rickett1977,Manchester2001}. As both of these effects worsen with decreasing frequency, and are more serious for faster pulsars, where small time delays to scattering, or due to dispersion too strong to correct across finite width channels, can lead to smear pulses out by more than the pulse period timing-based searches for distant pulsars are easier at higher frequencies, but where pulsars are dimmer due to their steep spectral index \citep{Bates2013}.
The amount of diffuse material along the Earth's line of sight also greatly increases toward the Galactic Centre, and accordingly the complications from high dispersion measure likely render many pulsars in this region, and certainly the vast majority of MSPs due to their very fast spin periods, effectively invisible to pulsar searches so far. Additional difficulties are present if the pulsar is a close binary, where either the orbital acceleration \citep{Johnston1991} or eclipsing from the companion or shed material \citep{Fruchter1988} can obscure the timing search.
To date, there have been a rather small number of pulsars found in the inner few degrees of the dense Galactic Centre with timing-based pulsar searches (e.g., $\sim$70 within a 4$^{\circ}$ radius, of which only 3 are MSPs with spin periods less than 10~ms in the Australia Telescope National Facility Pulsar Catalog (ATNFPC, \citep{Manchester2005}) -- \citealt{Deneva2009,Suresh2022}), despite more than 3000 pulsars discovered in total \citep{Manchester2005}.

There is another method that can search for evidence of pulsar populations on large scales, which utilizes imaging rather than timing observations. With imaging, spectral indices can be obtained, and the pulsars' very steep radio spectral indices set them apart from most other source classes (high-redshift accreting supermassive black holes of active galactic nuclei (AGN) \citealt{DeBreuck2000} and randomly scintillated sources \citep{Rickett1990} can also display very steep spectra). 
Imaging does not suffer from propagation effects like timing-based searches for pulsars do, and moreover imaging at lower frequencies is comparatively easier than timing-based pulsar searches
in order to observe brighter flux densities from the steep spectrum of any potential pulsars (regardless of spin period) in the field of view. 
Steep spectrum point sources are viable pulsar candidates \citep{Bhakta2017,Maan2018,Hyman2021}. Although a steep spectrum is not enough to confirm a particular source as a pulsar, when a strong statistical excess of steep-spectrum point sources is seen in a region, a comfortable conclusion may be drawn that most of the sources should indeed be pulsars (e.g., an observational over-density of pulsars in the Galactic Plane compared to higher latitudes, $\sim$60\% of pulsars within $5^{\circ}$ in the ATNFPC). Though there have also been populations unidentified radio point sources found previously, such sources are usually not as steep spectrum as pulsars, and are much brighter than expected for pulsars near the Galactic centre as well \citep[e.g.,][]{Crawford2000,Crawford2021,Wang2021}. This suggestive evidence from imaging and spectral indices is usable on the larger scales of surveys to draw general conclusions about the overall presence of pulsars in a region, and may currently be the most reliable way to uncover evidence of many of the pulsars in regions that are very difficult to observe with timing-based pulsar searches, such as the Galactic Centre and Bulge.  

Usually born from the core collapse of a massive star \citep{Burrows1986}, neutron stars begin their lives in the Galactic Plane, but do not necessarily stay there. Neutron stars may receive velocity kicks when they are born \citep{Blaauw1961,LyneLormier1994,Wheeler2000}, significantly increasing the magnitude of their spatial velocity compared to their progenitor stars, and applied in a randomized direction to their former orbital trajectory. 
Spatial velocities of pulsars resulting from these kicks 
have been observed to be potentially quite high at up to $\sim$1000~km~s$^{-1}$ \citep{Arzoumanian2002,Hobbs2005}, and lead to some fraction of pulsars able to either escape the inner Galaxy's gravitational potential, or enter an eccentric Galactic orbit and travel through the Bulge. This leads to the theory that velocity kicks should work to disperse pulsars into the Galactic Bulge \citep{BoodramHeinke2022}, with estimations of their numbers residing in the thousands \citep{Gonthier2018}.

Another process to consider that contributes to the pulsar population is the tidal breakup of former globular clusters (GCs), which inspiral toward the Galactic centre due to dynamical friction \citep{Tremaine1975}. This process is needed to explain the current distribution of GCs. The tidal breakup of GCs has been predicted to deposit their pulsars in the central regions of galaxies \citep{PhinneyKulkarni1994,Gnedin2014}, which alone should lead to the existence of a significant number of pulsars near the Galactic Centre (e.g., $\sim$1000 within 3$^{\circ}$ \citep{Brandt2015}).
There is also an unresolved gamma ray excess toward the centre of the Galaxy, called the Galactic Centre Excess (GCE) \citep{Acero2015}. A modelled population of $\sim$10$^{3-5}$ millisecond pulsars is able to at least partially reconcile the GCE \cite[e.g.][]{Brandt2015,Bartels2016,Gautam2021} (another explanation for the GCE is dark matter annihilation \citep{HooperGoodenough2011}).  

Overall, it is reasonable to suspect that there must be many more pulsars inhabiting the regions near the centre of the Galaxy in addition to the rather small number currently known, and imaging may be able to reveal evidence of their truer numbers. The population and distribution of pulsars have implications for theories of velocity kicks, globular cluster evolution, and the GCE, in addition to the evolution of pulsars themselves and the pathways to create them, as from massive stars with supernovae or direct collapse of white dwarf stars due to accretion \citep{Nomoto1991}.

\subsection{X-ray binaries}

Black hole X-ray binaries (BHXBs) are systems in which a black hole accretes material from a companion star. Almost all BHXBs spend long periods in a dim quiescent state, and occasionally undergo bright optical and X-ray outbursts which have been used to identify the vast majority of systems known to date \citep{Corral-Santana2016,Tetarenko2016WATCHDOG}.
BHXBs have been known from observational history to spend up to decades in quiescence between outbursts (e.g., a single known outburst or 4-5 decades between observed outbursts in some systems noted in \citet{Maccarone2022}), and it is likely that there is an even larger population of systems that are quiet for much longer, spending centuries or millennia in this inactive state. These long quiescence-period systems will never have shown an outburst in the era of X-ray astronomy, but their existence is required to reconcile the predictions of large numbers of Galactic BHXBs ($\sim$10$^{4}$) from 
evolutionary modeling \citep{PortegiesZwart1997,Jonker2011}, and the implied population may be even larger, based on a nearby radio-selected object in the foreground of M15 \citep{2016ApJ...825...10T}. If there are indeed a large number of systems that have been in quiescence for the past few decades and that will remain inactive for the foreseeable future, their detection and identification must be done without utilizing an outburst.

Quiescent systems do emit faintly in X-rays from their accretion disks and can be bright enough to detect with X-ray observatories, but another identification problem exists: in optical and in X-rays, quiescent BHXBs appear very similarly to cataclysmic variables (CVs), where a white dwarf accretes matter from a low-mass companion star \citep{Patterson1984}. CVs are vastly more numerous than BHXBs \citep[e.g., 1920 known CVs in][]{Abril2020}, despite the CVs typically being more nearby than the known X-ray binaries, so the natural assumption is that any system that appears as such is a CV rather than a quiescent BHXB (qBHXB). It is possible to see the  signature of the hot white dwarf in ultraviolet spectra, but this is restricted to sources with low extinction, and is not widely applicable for  sources outside the solar neighbourhood. Additional methods utilizing optical spectra with radial velocities and line widths \citep{Casares2016,Casares2018} are also possible with targeted follow-up observations. However, radio emission may also break the degeneracy: qBHXBs produce radio emission via jets \citep{Fender2004} which can be used to indicate that a particular system is a qBHXB rather than a CV \citep{Maccarone2005}. Some CVs have been observed to produce radio emission from a variety of mechanisms including jets and cyclotron masers, but relatively few CVs reach radio/X-ray ratios comparable to qBHXBs \citep{Mason&Gray2007,Coppejans2015,Ridder2023}, so a rough distance estimate with a radio counterpart may suggest that a particular source is more likely a BHXB instead. This X-ray/radio flux ratio, which few classes of objects are able to produce simultaneously, offers a method to identify candidate BHXBs independently of their outbursts \citep[e.g.][]{Zhao2020}.  Understanding the effectiveness of radio surveys for finding stellar mass black holes is also important for laying the groundwork for what can be done with future radio facilities like SKA and ngVLA \citep{fender2013,2018ASPC..517..711M}.

In addition to BHXBs, neutron star X-ray binaries (NSXBs) can also produce both X-ray and radio emission, similar to BHXBs. For a given X-ray luminosity, NSXBs are usually  significantly fainter in radio than BHXBs \citep{FenderKuulkers2001,vandenEijnden2021}, though some NSXBs also overlap with radio-bright BHXBs \citep{Panurach21}, such as the rare subclass of transitional millisecond pulsars \citep{Deller2015}. Overall, for NSXBs and BHXBs, radio observations may  aid in distinguishing between two types of similar sources, especially when data from other wavelengths does not allow for the nature of the source to be determined.

Overall, XRBs represent some of the last stages of binary and stellar evolution, but their observed numbers are currently much lower than models have predicted. Discovering more of these systems, which is difficult due to their primarily transient nature, will aid in population and evolution modelling by providing a better foundation for their properties such as their overall numbers, companion types, orbital periods, and distribution throughout the Galaxy \citep{Postnov2006}.

\begin{table*}
	\centering
	\setlength{\tabcolsep}{10pt}
	\begin{tabular}{llllc}
		\hline
		Catalogue & Full Name & Regime & Sens. | ang. res.$^{a}$ & Reference \\
		\hline\hline
		VLASS & VLA Sky Survey (Epoch 1.1) & 3~GHz & 0.6~mJy | 2" & 1 \\
		RACS & Rapid ASKAP Continuum Survey (low frequency) & 887~MHz & 1.5~mJy | 30" & 2 \\
		NVSS & NRAO VLA Sky Survey & 1.4~GHz & 3.8~mJy | 45" & 3 \\
        TGSS & TIFR GMRT Sky Survey & 150~MHz & 30~mJy | 25" & 4 \\
		\hline
        {\it Gaia} & {\it Gaia} & $\sim$800~nm & 21~mag | -- & 5 \\
		MIPSGAL & Micron Point Source Galactic Plane Survey & 24~$\mu$m & 3.3~mJy | 6" & 6 \\
		GLIMPSE & Galactic Legacy Infrared Mid-Plane Survey Extraordinaire & 3.6, 4.5, 5.8, 8.0~$\mu$m & 0.4~mJy | $\sim$2" & 7 \\
		WISE & Wide-field Infrared Survey Explorer & 3.4, 4.6, 12, 22~$\mu$m & 0.1--6~mJy | 6--12" & 8 \\
		\hline
		CXOGBS & {\it Chandra} X-ray Observatory Galactic Bulge Survey & 0.3--8~keV & $2.3 \times 10^-13 \mathrm{erg~s^{-1}~cm^{-2}}$ | 1" & 9 \\
        {\it Fermi} & Fermi Gamma-ray Space Telescope & 50~MeV--1~TeV & $\sim10^{-12}~\mathrm{erg~s^{-1}~cm^{-2}}$ | $\sim$1$^{\circ}$ & 10 \\
		\hline
		ATNFPC & Australia Telescope National Facility Pulsar Catalog & Pulsars (0.1 -- 10~GHz) & -- | -- & 11 \\
		\hline
        VLAGBS & Very Large Array Galactic Bulge Survey & 1--2~GHz & 0.1~mJy | 1" & this work \\
		\hline
	\end{tabular}

 \caption{List of catalogues referenced in this paper or used for matching, either with Vizier \citep{Vizier2000} or Astroquery \citep{Astroquery2019}. Column$^{a}$ is the typical values for sensitivity at the 5$\sigma$ (5~$\times$~RMS) level | angular resolution. \\
	1. \protect\url{https://science.nrao.edu/vlass} ---  \citet{Lacy2020} \\
	2. \protect\url{https://research.csiro.au/racs/home/survey} --- \citet{McConnell2020} \\
	3. \protect\url{https://www.cv.nrao.edu/nvss} --- \citet{Condon1998} \\
    4. \protect\url{https://tgssadr.strw.leidenuniv.nl/doku.php} --- \citet{TGSS2017} \\
    5. \protect\url{https://www.cosmos.esa.int/web/gaia-users/archive/gdr3-documentation} --- \citet{Gaia2023} \\
	6. \protect\url{http://mipsgal.ipac.caltech.edu} --- \citet{Carey2009}  \\
	7. \protect\url{http://www.astro.wisc.edu/glimpse} --- \citet{Benjamin2003} \\
	8. \protect\url{https://www.nasa.gov/mission\_pages/WISE/main/index.html} --- \citet{Wright2010}  \\
	9. \citet{Jonker2011}  \\
    10. \protect\url{https://fermi.gsfc.nasa.gov} --- \citet{Atwood2009Fermi} \\
    11. \protect\url{https://www.atnf.csiro.au/research/pulsar/psrcat} --- \citet{Manchester2005}
}
\label{table:Catalogs}
\end{table*}

\subsection{Other stellar radio phenomena}

Other sources such as young stellar objects and planetary nebulae may be lost in the crowded infrared and optical wavelengths of the Centre and Bulge, especially if they are faint, but radio counterparts to such sources may allow them to stand out and be more easily identified. For example, planetary nebulae are extended sources that can be resolved with radio interferometry, and their nebular emission has flat or inverted spectral indices \citep{Wright1975}. A radio source that fits this description can then be compared to infrared catalogs and data to determine if there is a coincident infrared counterpart that follows their typical spectral energy distributions that peak in the mid-infrared and extend to the radio \citep{Zhang1991,Anderson2012}. The combination of a radio source's flux, spectral index, and morphology from imaging allows them to be effective informants when classifying the radio sources themselves, or when used as counterparts with other wavelengths. Even the mere presence of a radio counterpart can significantly narrow down potential classifications due to the limited range of object classes that are detected in radio. Surveys for objects like planetary nebulae in the radio avoid exctinction biases of other approaches; the relatively small number of new planetary nebula candidates here is likely a result of the shallowness of our survey, but still highlights a path forward for finding more objects through this technique.

With the higher stellar densities toward the Galactic Centre, a relatively small survey area is able to observe a large number of Galactic objects. The Galactic Bulge alone contains $\sim$5\% of the stellar mass of the Milky Way \citep{Valenti2016}. As much of this region lies above and below the Plane, observations of it avoid the most severe extinction issues, while still viewing one of the most spatially dense areas of the Galaxy. This makes the Galactic Bulge a good choice for carrying out multiwavelength surveys focused on detecting Galactic sources.

\subsection{Plan of the paper}

\begin{figure}
    \includegraphics[width=0.99\linewidth]{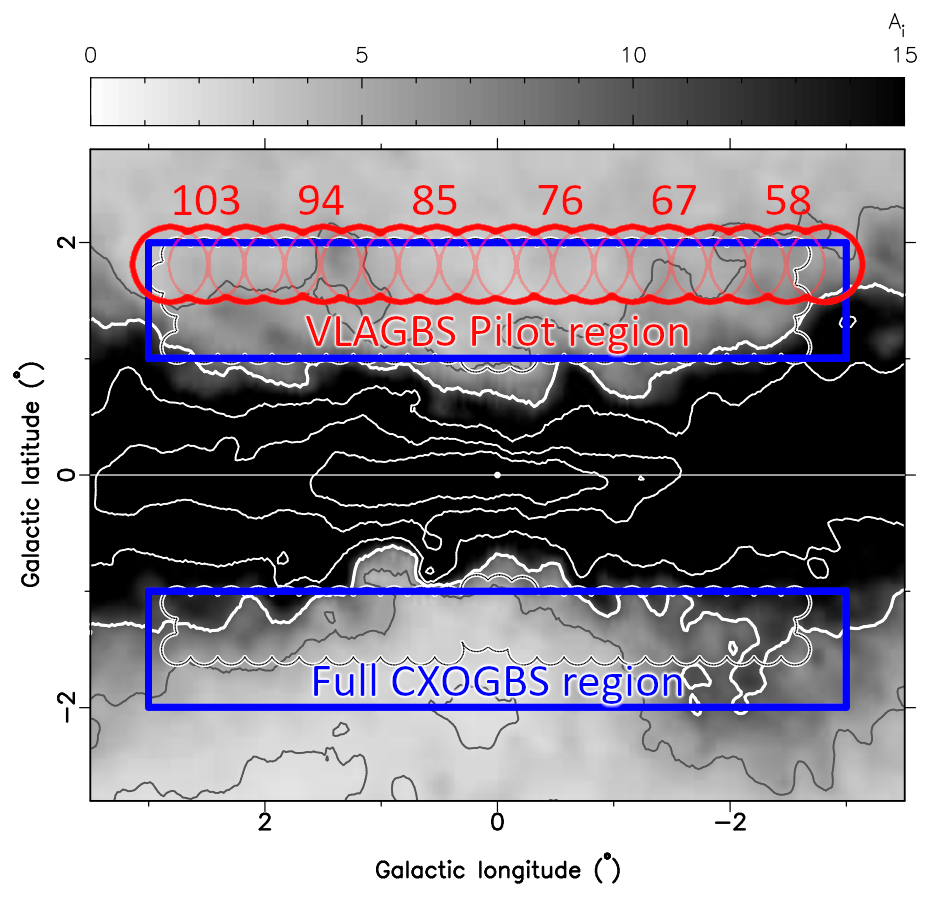}
	\caption{Galactic extinction map background with the full CXOGBS region indicated in blue rectangles, and this VLAGBS pilot survey region indicated in red, where the red circles are the individual 18 fields that comprise the survey (every third field is labelled by the field number from Table~\ref{tab:tableFields}). Figure adapted from \citet{Jonker2011}. 
	\label{fig:map}}
\end{figure}

We present the results of the Very Large Array Galactic Bulge Survey (VLAGBS) pilot, a radio imaging survey at 1--2~GHz in the Galactic Bulge region covering $\sim$3 square degrees with a maximum sensitivity of $\sim$0.1~mJy and angular resolution of 1.1", as a complement to the 
{\it Chandra}
Galactic Bulge Survey 
\citep[CXOGBS,][]{Jonker2011,Jonker2014}, and to the numerous public infrared surveys (e.g., those listed in Table~\ref{table:Catalogs}) covering this region of the Galaxy. The VLAGBS covers one portion ($\sim$6$^{\circ}$ by 0.5$^{\circ}$; one row of pointings aligned such that the edge of each field of view meets the center of its neighboring fields, see Figure~\ref{fig:map}) of the total GBS region (12 square degrees) for a first look at the Bulge with such sensitivity and resolution in radio, higher in both respects than current all-sky radio surveys. With the X-ray sources already identified and, when possible, classified with optical or other wavelength counterparts, the addition of a radio counterpart may help to classify or re-classify objects found in the CXOGBS. 
Spectral indices were calculated for each source within the 1--2~GHz frequency band to provide more information on the possible mechanism of the radio emission. In total we find 1617 unique radio sources, 25 of which have X-ray counterparts from the CXOGBS, and about 100 of which are steep spectrum point sources ($\alpha \lesssim -1.4$) and thus viable pulsar candidates. We identify a handful of particularly notable sources as well as additional interesting ones, and discuss these individual sources and the overall results. The data and analysis are described in Section~\ref{Data}, overall results of the VLAGBS and matching to multiwavelength catalogs are discussed in Section~\ref{Results}, particular notable sources in Section~\ref{Notable Sources}, and conclusions in Section~\ref{Conclusions}.

\section{Data}\label{Data}

\begin{table*}
	\centering
	\setlength{\tabcolsep}{20pt}
	\begin{tabular}{cccc}
		\hline
		  &   & Field Centre \\
		Obs. Date & Field number & J2000 & Galactic: l, b ($^{\circ}$)\\
		\hline\hline
		  & 64 & 17:34:02.161 --29.30.50.639 & --1.83256, 1.83288\\
		11 July 2015 & 67 & 17:34:51.850 --29.14.02.357 & --1.49924, 1.83288\\
		  & 70 & 17:35:41.253 --28.57.12.813 & --1.16592, 1.83288\\
		\hline
		  & 82 & 17:38:56.100 --27.49.42.356 & 0.16738, 1.83290\\
		23 July 2015 & 85 & 17:39:44.139 --27.32.46.799 & 0.50070, 1.83290\\
		  & 88 & 17:40:31.918 --27.15.50.097 & 0.83402, 1.83291\\
		\hline
		  & 91 & 17:41:19.441 --26.58.52.238 & 1.16736, 1.83291\\
		24 July 2015 & 94 & 17:42:06.711 --26.41.53.300 & 1.50068, 1.83291\\
		  & 97 & 17:42:53.733 --26.24.53.272 & 1.83400, 1.83292\\
		\hline
		  & 55 & 17:31:31.335 --30.21.07.738 & --2.83253, 1.83286\\
		25 July 2015 & 58 & 17:32:21.907 --30.04.23.366 & --2.49921, 1.83287\\
		  & 61 & 17:33:12.180 --29.47.37.669 & --2.16589, 1.83287\\
		\hline
		  & 100 & 17:43:40.511 --26.07.52.140 & 2.16733, 1.83292\\
		26 July 2015 & 103 & 17:44:27.047 --25.50.49.982 & 2.50065, 1.83293\\
		  & 106 & 17:45:13.345 --25.33.46.786 & 2.83397, 1.83293\\
		\hline
		  & 73 & 17:36:30.377 --28.40.21.997 & --0.83259, 1.83289\\
		27 July 2015 & 76 & 17:37:19.222 --28.23.29.989 & --0.49927, 1.83289\\
		  & 79 & 17:38:07.794 --28.06.36.779 & --0.16595, 1.83289\\
		\hline
	\end{tabular}
	\caption{List of the observations of VLAGBS and pointings of the fields.
	\\ 
    }
	\label{tab:tableFields}
\end{table*}

\subsection{Observations and data reduction}

The VLAGBS was carried out with the Very Large Array (VLA) over six days in July 2015, with a total of 18 fields/pointings (Project Code: 15A--073). It covers a portion of the Galactic Bulge region, spanning the Galactic coordinate range of $-3^{\circ}<l<+3^{\circ}$ and $+1.6^{\circ}<b<+2.1^{\circ}$, for a total of about 3 square degrees. Observation dates and field pointings are listed in Table~\ref{tab:tableFields}. The observations were done in A-configuration and L-band (36~km maximum baseline, 0.68~km minimum; 1--2~GHz, 16 spectral windows), for an average angular resolution of 1.1"$\pm$0.1" for the minor axis ($\sim$East-West), and 2.7"$\pm$0.3" for the major axis ($\sim$North-South). Each field was observed for roughly 20.5 minutes in total over four scans. The phase calibrator was J1751--2524 and the flux calibrator was 3C286 (J1331+305) for all observations. Data were first processed through the VLA calibration pipeline in Common Astronomy Software Applications (CASA) \citep{McMullin2007}. Data were then flagged for remaining radio frequency interference (RFI) with \verb+mode=’tfcrop’+ and \verb+‘rflag’+, and visually inspected for remaining bad data and flagged manually. An image of each field of the full 1--2~GHz bandwidth was manually cleaned via CASA's \verb+tclean+ task with parameters of \verb+deconvolver=mtmfs+, \verb+nterms=2+, \verb+gridder=widefield+, and \verb+robust=0.5+, including a phase self-calibration (\verb+solint=inf+) to reduce artifacts. The data for each field were also divided into four subband images based on remaining spectral windows after flagging, and each of these images was cleaned similarly. The subband images were used to calculate spectral indices.

\subsection{Source selection}

A significant fraction of sources in any field are fainter than the artifacts surrounding the brightest few, and many sources are extended in morphology, e.g., AGN and planetary nebulae: these factors can complicate source identification. We tested the Python Blob Detection and Source Finder software (PyBDSF, \citealt{Mohan2015}; 4$\sigma$ island threshold, 4$\sigma$ peak, source size not fixed to beam), as well as CASA's \verb+image.findsources+ function to detect sources in the full bandwidth images. Both methods resulted in numerous spurious noise detections (e.g., on artifacts emanating from bright sources that are not well-shaped to the synthesized beam), as well as non-detections of isolated $>5\sigma$ point sources. To most accurately identify real sources in the survey, the full bandwidth images without primary beam correction were visually inspected for sources with the PyBDSF results overlayed. To select sources manually, the cleaned image of the full 1--2~GHz was used, as this image has the lowest background level. Contours were used to guide the source selection: the lowest contour level was set such that peak fluxes of roughly 4$\sigma$ were enclosed with a contour, adjusted accordingly as the background root mean square (RMS) varied over the image. Selected sources were boxed with the polygon selection tool, and the total, peak, and local RMS (surrounding $\sim$1' diameter region, real sources excluded) flux densities were recorded. 

To correct for the primary beam, these values were divided by the primary beam response function (Equation 4 from EVLA Memo \#195\footnote{\protect\url{https://library.nrao.edu/public/memos/evla/EVLAM_195.pdf}}), dependent on the distance of the source from the centre of the field. The response function used was verified by comparing flux values of sources in primary beam corrected images (using \verb+pbcor=True+ in \verb+tclean+; the pbcor and non-pbcor imaging were done separately). Using the non-primary beam corrected images to select sources was chosen in order to keep the image background minimized and constant over the entire field (as the primary beam correction effectively amplifies the value of the background dependent on distance from the centre of the field, by up to a factor of $\sim$5 at the edge), making the identification of real sources easier and more reliable to do both manually and with automated methods. Since the largest angular scale to which our observations are sensitive ($\sim$36") is much less than the size of the fields (radius of $\sim$20'), using one primary beam response value for extended sources is a small discrepancy that is much less than other causes of uncertainty. All final source entries in the full catalog have a peak signal-to-noise ratio (S/N) $\geq 4$, though we do not claim completeness down to $4\sigma$. We restrict our analysis to sources with S/N~$\geq 5$ in Section~\ref{Results}, for which we are confident in a very high completeness value.

The regions of all sources in each field found in the full bandwidth images were saved and loaded onto the cleaned images of the subbands. Every source was revisited in each of the four subbands. If the source was apparent by a similar method of using contours to judge the brightness of the source relative to the level of the background noise, a new polygon region was drawn around the source to compensate for possible small position shifts and for the different resolutions of the subband images. Again the flux density and local background RMS were recorded for each source. If the source was not apparent at the location found in the full bandwidth, the background RMS was used to set an upper limit for a non-detection. Some sources that were located near the edge of the field fall off of the images at higher frequencies. The full bandwidth images have a background RMS of $\sim$20-30~$\mu$Jy in the centre, and the subband images have a central RMS of $\sim$35-150~$\mu$Jy, with the variation primarily due to spectral window flagging; a table of all fields and subband RMS values is provided in Table~\ref{tab:fieldRMS}.

All sources have major and minor axes measurements, and all were visually inspected. For sources that were detected by PyBDSF, it was checked that PyBDSF accurately captured the shape of the source. For sources correctly detected by PyBDSF, those measurements were used. For sources that PyBDSF either did not detect or did not correctly capture, measurements of the axes were made by hand for extended sources, and for point sources the synthesized beam parameters for the field were used.

\subsection{Spectral index calculation}

The spectral index for each source was calculated using the flux densities in the subbands, performing a powerlaw fit where the spectral index $\alpha$ is defined by $S_{\nu} \propto \nu^{\alpha}$, and $S_{\nu}$ is the flux density at the central frequency of the image (flux densities are weighted by their respective error value, set to the RMS value + 5\% of the total flux density; 5\% of the total flux density empirically accounts for potential under- and over-drawing of the source's region box). There are many faint sources that have either one or no detections in the subbands due to the decreased S/N from a higher background level, and these sources do not have a spectral index calculated. Sources that are extended may have calculated spectral indices that are steeper than their true values, either due to the loss of extended emission to decreased flux/beam, or over-resolving of structure in the higher frequency subbands due to decreased largest angular scale. Subband occurrences of extended sources that clearly lost a significant amount of structure were not assigned a flux density value for those subbands.

\subsection{Positional errors} \label{section:positionError}

As this survey is at 1--2~GHz, high resolution, and was observed with the VLA for the Galactic Bulge which is at a low elevation, there was a potential for the initially measured source positions to be slightly inaccurate (e.g., offsets introduced during phase calibration or due to far field correction discrepancies, which have been noted in past radio survey projects as well). We investigated the quality of positions of sources in the survey with two methods. Due to field overlap in our survey, some sources appear in two fields. These $\sim$300 sources were identified and the positions of the two appearances compared, using the entry which was closer to the centre of its respective field as the base position. 
The other check was made against VLASS 1.1 (with the updated positions from VLASS Project Memo \#14\footnote{\protect\url{https://library.nrao.edu/public/memos/vla/vlass/VLASS_014.pdf}}), though we only find $\sim$100 point sources with a VLASS counterpart. We also attempted to match positions to {\it Gaia} and to quasar catalogs, but found that there are too many optical stars that are unassociated to the radio sources to recover any match pattern, and that there were no matched quasars to the radio sources in this region of the sky.

We do not find a positional correction that improves both the field overlap discrepancies and VLASS matched positions, so we do not attempt to calibrate out any potential systematic positional errors, and note that the mean/median/standard deviation in arcseconds of point source positional discrepancies are for 108 VLASS matches less than 1'' in radius (values in arcseconds): overall 0.44/0.43/0.18; RA 0.15/0.11/0.14; and Dec 0.38/0.37/0.17; and for 273 field overlap sources: overall 0.31/0.29/0.19; RA 0.19/0.16/0.15; and Dec 0.21/0.16/0.18. From these results we use an overall 0.5" positional error value for the radio sources when matching the VLAGBS sources to other catalogues.

\subsection{Potential spectral index errors} \label{sec:specIndErr}

Sources that were both faint and close to the edge of the (full bandwidth) field were found to have a much higher probability of having very inverted or steep spectral indices as calculated from the subband fluxes. 
The primary beam correction amplifies the values of the effective background and thus the flux fluctuation amplitudes, and may 
cause the resulting spectral index to vary wildly. We found a tendency for this effect based on the spectral indices with the combination of low flux and high distance from the centre of the field (see Fig~\ref{fig:spedIndIssue} for plot of this effect). A similar effect was noted in \citet{Smolcic2017}, who found that spectral indices tended to be steeper on average further from the pointing centre. 

Because of this finding, and the reduced primary beam response further from the centre of the field (or, equivalently, an increase in the background RMS of primary beam-corrected images and data further from the centre), the sources with duplicated entries in the catalog from field overlap have the duplicate marker applied to the entry that is further from the centre of its respective field image. Marking the duplicates in this way results in the primary entry generally being the one with lower flux and spectral index errors while using an objective criterion. For the analysis of sources that follows in the remainder of the manuscript, we only consider the primary source entries.

We also found two sources that are extended in morphology, are significantly detected at S/N~$>20$, have very steep spectra of $\alpha<-3$, and that, if real, would be classified as transients due to their lack of a detection in RACS. On close visual inspection these sources lie at the intersection of faint lines of artifacts from sources either on- or off-field. We believe these two sources to be spurious and randomly amplified manifestations of artifact interference. We removed these two sources from the final $\geq5\sigma$ catalog, and images of them are provided in Figure~\ref{image:artifactAmpSources}.

\begin{table*}
	\centering
	\setlength{\tabcolsep}{10pt}
	\begin{tabular}{ccccc|cc}
		\hline
		 ATNFPC &  &  &  &  & VLAGBS & \\
		 Pulsar & S$_{\mathrm{1.4~GHz}}$ (mJy) &  P$_{\mathrm{spin}}$ (s) & DM & ATNFPC Ref. &  S$_{\mathrm{1.5~GHz}}$ (mJy) & $\alpha$ \\
		\hline\hline
		J1734--2859 & 0.13 & 0.30 & 314 & \citet{Cameron2020} & 3$\sigma<0.19$ & -- \\ 
		J1736--2843 & 0.43 & 6.44 & 331 & \citet{Hobbs2004} & $0.39\pm0.05$ & $-1.47\pm0.65$\\ 
		J1736--2819 & 0.16  & 1.59 & 261 & \citet{Hobbs2004}& 3$\sigma<0.13$ & -- \\ 
		J1738--2736 & 0.17 & 0.63 & 324 & \citet{Ng2015} & 3$\sigma<0.12$ & -- \\ 
		J1741--2719 & 0.20 & 0.34 & 362 & \citet{Lorimer2006} & $0.25\pm0.05$ & -- \\ 
		\hline
	\end{tabular}
	\caption{Comparison of cataloged ATNFPC pulsars located in the survey fields.  Flux densities are in units of mJy. Upper limits are $3\sigma$. We only recover two of the five ATNF pulsars 
 in our survey. All listed ATNFPC values of the 1.4GHz flux, spin period, and dispersion measure (cm$^{-3}$pc) are from the same reference.}
	\label{tab:tableATNF}
\end{table*}

\section{Results} \label{Results}

There are 1617 unique sources with S/N~$\geq$~5, about 8 times the number of VLASS sources (198) in the survey region, and 4 times the number of RACS sources (429). There are 1331 point or near-point sources defined by total/peak flux ratios $\leq 1.5$, and 140 sources which we note as likely or certainly associated to AGN emission, i.e. double lobes or core and hotspot morphologies, as well as 31 known planetary nebulae. For the following, we consider only the primary entry for sources as described above.

Each source has a spectral index calculated within the 1--2~GHz observation bandwidth if there are at least two subband detections: there are 969 unique sources with spectral indices calculated. Of the 648 without a spectral index, 604 are point sources that are faint and not detected in at least two subbands. The remaining extended sources may either be faint or have suffered from over-resolving. 

In the following subsections we discuss the steep spectrum point source population in the context of pulsars, and match the VLAGBS catalog to various multiwavelength catalogs. We also provide an estimate for the number of radio AGN expected to be present in the survey region.

\subsection{Steep spectrum point sources}

There are 169 point-like sources with spectral indices steeper than --1, and 91 of these are steeper than --1.4 at their central spectral index values. 
One possibility is that these are predominantly high redshift galaxies which have steep radio spectra. \citet{DeBreuck2000} find that 0.5\% of radio sources in their joint WENSS/NVSS sample have radio indices $\alpha < -1.3$; applying this naively to our catalogue (assuming that most of our radio sources are extragalactic, and that the fraction that are steep-spectra high-z radio galaxies is constant with flux) would suggest that $\sim$ 8 of our very steep radio sources are likely to be high-redshift radio galaxies, although the number may be a bit smaller given the higher angular resolution of our survey relative to WENSS and NVSS, which may resolve a larger fraction of these objects. Most pulsars have very steep spectral indices, with $< \alpha > $ $ = -1.4$ \citep{Bates2013}, thus these sources may predominantly be pulsars instead. Indeed, we do find that this survey contains an excess of steep spectrum sources $\alpha \lesssim -1$ compared to extragalactic surveys such as \citet{Smolcic2017}.

About 100 pulsars were expected to be in this survey region at the given sensitivity based on the modeling in \citet{Jonker2011}, aligning very well with our results.
By contrast, using the ATNF pulsar catalog \citep{Manchester2005}, there are only 5 pulsars currently cataloged within the survey region (Table~\ref{tab:tableATNF}).
The observed excess of steep spectrum sources is suggestive evidence that there is a significant pulsar population near the Galactic Centre that has evaded detection in timing-based pulsar surveys so far, and that supports various evolutionary theories that produce large numbers of pulsars in the Galaxy's history and may also explain the GCE. The restriction of parameter space of timing-based searches for pulsars in regions of high dispersion measure, especially for millisecond spin periods, is presumably the reason this sizable missing pulsar population has not been found to date despite large scale timing-based pulsar survey campaigns \citep[e.g.][]{Manchester2005,Keith2010,Chennamangalam14,Macquart15}. Our result illustrates that imaging could be a productive effort to find evidence of large populations of pulsars, a method that has also been used successively in the past with followup timing observations \citep{CordesLazio,Bhakta2017,Hyman2019}. However, again, only the detection of radio pulsations can confirm a pulsar, and thus any source found via imaging cannot be confirmed using spectral index alone. 
Most of our steep spectrum point sources have flux densities of less than 0.5~mJy at 1.5~GHz; fast pulsars with such fluxes are not expected to be detected in this region of the Galaxy in pulsar surveys based on their parameter spaces.

\begin{table*}
	\centering
	\setlength{\tabcolsep}{4pt}
	\begin{tabular}{lccccccl}
		\hline
		 CXOGBS & 0.3-8~keV flux & VLAGBS & S$_{\mathrm{1.5GHz}}$ & $\alpha$ & log($\mathrm{L_{R}}/\mathrm{L_{X}}$) & RS & Notes\\
		 source & 10$^{-14}$erg s$^{-1}$ [counts] & source & mJy &  &  & \\
		\hline\hline
		CX40 & 27.16 [35] & J174404.3--260925 & 17.58$\pm$0.91 & 0.36$\pm$0.14 & --12.2 & RNV$^{\dagger}$ & AGN \\ 
		CX49 & 23.28 [30] & J173146.8--300308 & 62.40$\pm$3.15 & 0.23$\pm$0.13 & --11.6 & RNV$^{\dagger}$ & AGN \\ 
		CX51 & 22.50 [29] & J174000.6--274859 & 0.97$\pm$0.09 & --2.31$\pm$0.54 & --13.4 & -- & *YSO/MSP/AGN? \\ 
		\arrayrulecolor{lightgray}\hline
		CX63 & 20.17 [26] & J173411.3--293117 & 0.27$\pm$0.05 & --0.80$\pm$0.91 & --13.9 & -- & *CV, now BHXB candidate \\ 
		CX68 & 18.62 [24] & J173204.6--295220 & 2.61$\pm$0.17 & 0.45$\pm$0.26 & --12.9 & RV & \\ 
		CX78 & 17.85 [23] & J173354.1--292138 & 0.36$\pm$0.06 & 0.27$\pm$0.78 & --13.7 & -- & \\ 
		\hline
		CX101 & 14.74 [19] & J173353.1--300131 & 0.80$\pm$0.10 & --0.20$\pm$1.51 & --13.3 & -- & \\ 
		CX233 & 7.76 [10] & J174206.1--264115 & 6.31$\pm$0.34 & --0.79$\pm$0.14 & --12.1 & RNV$^{\dagger}$ & AGN \\ 
        CX273 & 6.98 [9] & J174239.6--262655 & 1.29$\pm$0.087 & --0.82$\pm$0.21 & --12.7 & RV$^{a}$ & $^{b}$3.4" match radius, large X-ray position error \\ 
		CX278 & 6.98 [9] & J174044.5--265913 & 0.26$\pm$0.04 & -- & --13.4 &  &  \\ 
		\hline
		CX293 & 6.98 [9] & J174000.6--274816 & 174.77$\pm$8.78 & --0.92$\pm$0.14 & --10.6 & RNV$^{\dagger}$ & AGN \\ 
		CX330 & 6.21 [8] & J173643.8--282121 & 0.73$\pm$0.07 & 0.87$\pm$0.38 & --12.9 & -- & *Infrared transient, YSO? \\ 
		CX392 & 5.43 [7] & J173551.3--285646 & 0.16$\pm$0.03 & -- & --13.5 & -- &  \\ 
		\hline
		CX455 & 4.66 [6] & J174442.9--254932 & 0.43$\pm$0.05 & 0.07$\pm$0.52 & --13.0 & -- &  \\ 
		CX480 & 4.66 [6] & J173801.2--281351 & 0.25$\pm$0.04 & --1.56$\pm$0.83 & --13.3 & -- & Possible pulsar + M dwarf binary \\ 
		CX488 & 4.66 [6] & J173605.2--283238 & 122.31$\pm$6.15 & --1.25$\pm$0.13 & --10.6 & RNV$^{\dagger}$ & $^{b}$4.5" match radius, extended, likely AGN \\ 
		\hline
		CX498 & 4.66 [6] & J173419.6--294549 & 0.37$\pm$0.07 & -- & --13.1 & -- &  \\ 
		CX597 & 3.88 [5] & J174239.6--262655 & 1.08$\pm$0.10 & --0.82$\pm$0.21 & --12.6 & R &  \\ 
		CX639 & 3.88 [5] & J173242.3--295039 & 1.90$\pm$0.14 & --0.11$\pm$0.22 & --12.3 & RV & AGN \\ 
		\hline
		CX875 & 3.10 [4] & J173650.8--284539 & 0.15$\pm$0.04 & -- & --13.3 & -- & \\ 
		CX1146 & 2.33 [3] & J173832.3--280823 & 0.11$\pm$0.03 & -- & --13.3 & -- & \\ 
		CX1215 & 2.33 [3] & J173340.4--293332 & 0.16$\pm$0.04 & -- & --13.2 & -- & \\ 
		\hline
		CX1222 & 2.33 [3] & J173239.3--300701 & 0.17$\pm$0.04 & -- & --13.1 & -- & \\ 
		CXB163 & 4.66 [6] & J173228.3--302534 & 0.39$\pm$0.05 & 1.30$\pm$0.72 & --13.1 & RN$^{\dagger}$ & Likely AGN \\ 
		CXB290 & 3.10 [4] & J173138.5--302946 & 0.20$\pm$0.04 & --1.24$\pm$2.32 & --13.2 & -- & \\ 
		\arrayrulecolor{black}\hline
	\end{tabular}
	\caption{Matched CXOGBS X-ray sources with radio sources in our survey, of positions within 1.5" and most well within 1". The chance match rate within 2" is very low ($\sim$1\%). The RS column denotes whether the X-ray source has a radio counterpart in Radio Surveys of RACS (R), NVSS (N), or VLASS (V). Source$^{a}$, CX273, is not in the VLASS catalog, but is visually present in the images. Sources$^{b}$, CX273 and CX488, have larger match radii. Asterisk* sources are discussed further in Section~\ref{Notable Sources}. $^{\dagger}$ sources were previously reported as NVSS and CXOGBS matches in \citet{Maccarone2012}, or \citet{Jonker2014} for B163. The spectral index and log($\mathrm{L_{R}}/\mathrm{L_{X}}$) can aid in classifying sources; a table for these classifications is in \citet{Maccarone2012}. Sources with no notes are radio point sources with no additional information from counterparts. See Section~\ref{Results} for further discussion.}
	\label{table:CXOmatches}
\end{table*}

\begin{table*}
	\centering
	\setlength{\tabcolsep}{4pt}
	\begin{tabular}{llcccl}
		\hline
		 Catalog & Fermi & PowLaw & Assoc? & \# Radio & Notes \\
		\hline\hline
		4FGLb & J1731.6-3002 & 2.26 & A & 2 & Associated AGN \\
		4FGLb & J1733.2-2915 & 2.41 & U & 3 & Faint point sources \\
		4FGLb & J1733.5-2950 & 2.24 & U & 9 & One AGN, one steep spectrum point source$\mathrm{^{a}}$ \\
		\arrayrulecolor{lightgray}\hline
		4FGLb & J1734.0-2933 & 2.35 & U & 6 & Three 1-2~mJy point sources, one AGN \\
		4FGLb & J1734.6-2912 & 2.30 & U & 8 & One $\sim$20~mJy point source, two AGN \\
		4FGLb & J1734.0-2933 & 2.31 & U & 6 & Two 4-5~mJy point sources, one AGN \\
		\arrayrulecolor{lightgray}\hline
		3FGL & J1736.5-2839 & 2.50 & A & 25 & Associated supernova remnant; one AGN, multiple steep spectrum point sources \\
        4FGLb & J1737.1-2901 & 2.52 & U & 7 & One AGN, one steep spectrum point source \\
		4FGLb & J1737.2-2808 & 2.52 & U & 7 & Three 2-5~mJy point sources \\
        \arrayrulecolor{lightgray}\hline
		4FGLb & J1739.2-2717 & 2.51 & U & 6 & Faint point sources \\
		4FGLb & J1739.2-2732 & 2.46 & U & 3 & Faint point sources \\
		3FGL & J1740.5-2726  & 2.57 & U & 12 & One 1~mJy point source \\
        \arrayrulecolor{lightgray}\hline
		4FGLb & J1740.7-2640 & 2.48 & U & 1 & Potential tMSP$^{\dagger}$? \\
        4FGLb & J1741.6-2730 & 2.69 & U & 3 & Two AGN \\
		4FGLb & J1744.5-2612 & 2.35 & A & 1 & Associated to unclassified X-ray source (no radio counterpart) \\
        4FGLb & J1746.1-2541 & 2.56 & U & 2 & One AGN, one steep spectrum point source \\
        \arrayrulecolor{black}\hline
  
	\end{tabular}
	\caption{Matched {\it Fermi}-LAT sources in our survey region, selecting only the most recent catalog appearance for each source. Catalogs in the first column are: 3FGL \citep{Acero2015}; and the incremental 4FGL catalog denoted as 4FGLb \citep{Abdollahi2022}. The gamma ray power law is taken from the corresponding catalog noted in the first column, as well as the {\it Fermi} association status (A for associated, U for unassociated source). Notes include {\it Fermi} catalog notes, and notes from our radio survey. Faint point sources have either unknown spectral indices, or spectral indices that are consistent with synchrotron emission of about $-0.7$. $\mathrm{^{a}}$SSPS is a steep spectrum point source, where $\alpha < -1.4$. Other sources without notes in the region are faint point sources. Source $^{\dagger}$ is further discussed in Section~\ref{Notable Sources}.}
	\label{table:Fermi}
\end{table*}

\begin{table*}
    \setlength{\tabcolsep}{10pt}
	\begin{tabularx}{\linewidth}{@{}c X @{}}
    \begin{tabular}{ccccc}
	    \hline
	    Flux bin & Estimated & Catalogued & Cat. - Est. & Cat./Est. \\
        (mJy) &  Miller+13 & VLAGBS &  & \\
		\hline\hline
		0.2--0.3 & 348 & 279 & --69 & 0.8 \\
        0.3--0.4 & 147 & 158 & +11 & 1.07 \\
        0.4--0.8 & 222 & 252 & +30 & 1.14 \\
        \arrayrulecolor{lightgray}\hline
        0.8--1.6 & 148 & 167 & +19 & 1.13 \\
        1.6--3.2 & 65 & 104 & +39 & 1.6 \\
        3.2--6.4 & 83 & 73 & --10 & 0.88 \\
        \arrayrulecolor{lightgray}\hline
        6.4--12.8 & 28 & 31 & +3 & 1.11 \\
        12.8--25.6 & 28 & 32 & +4 & 1.14 \\
        25.6--51.2 & 28 & 8 & -20 & 0.29 \\
        \arrayrulecolor{lightgray}\hline
        51.2--102.4 & 19 & 9 & -10 & 0.47 \\
        > 102.4 & 16 & 0 & +7 & --- \\
		\arrayrulecolor{black}\hline
    \end{tabular}
    & 
    \includegraphics[width=0.9\linewidth, valign=c]{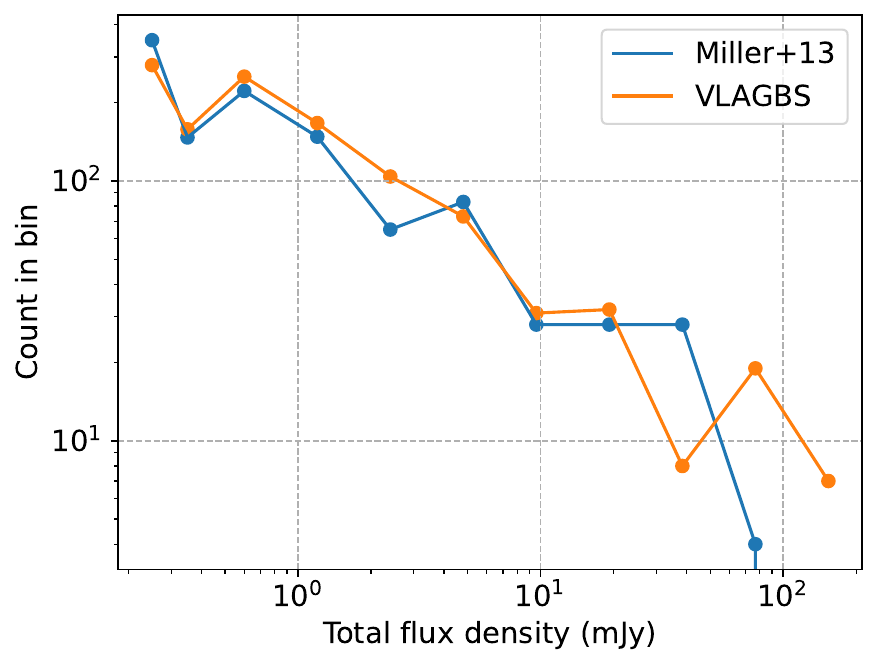}
    
	\end{tabularx}
	\caption{Estimates of the number of AGN expected in the survey region following the data of the VLA ECDF-S \citep{Miller2008,Miller2013} from Section~\ref{subsec:AGN}. 
	\label{table:AGN_est_Miller}.}
\end{table*}

\subsection{CXOGBS match}

We matched our radio catalog to the complementary {\it Chandra} Galactic Bulge Survey (CXOGBS) \citep{Jonker2011, Jonker2014}, defining the search radius as the position error of the X-ray source plus 0.5" to account for radio source position errors as noted in Section~\ref{section:positionError}. We use a combination of statistical methods and visual inspection of matching results and find a high match density and low chance match rate less than 2" in radius. In total we identify 23 strong CXOGBS matches and 2 additional notable potential matches of a larger match radius, which are all listed in Table~\ref{table:CXOmatches}.
We find radio counterparts to 6 either confirmed or candidate CXOGBS AGN, 1 compact binary, and 1 young stellar object, and the remaining 17 matches were unclassified in CXOGBS. A short summary of each of these sources follows. 

%
%

CX40 was confirmed as an AGN in \citet{Maccarone2012} with a spectrum and redshift. CX49 and CX233 were AGN candidates and are point sources in VLAGBS. CX293 is morphologically an FR-II radio AGN \citep{FanaroffRiley1974} in VLAGBS. The X-ray matches of CX639 and CXB163 coincide with point sources in VLAGBS, with two additional radio sources relatively close on opposite sides (45" and 15" from centre to end sources, respectively), equidistant and forming a linear line in total. This suggests that CX639 and CXB163 are also AGN, where the central radio source is the nucleus with the X-ray counterpart, and the two end radio sources are hotspots of otherwise unseen (possibly over-resolved) jets.

CX63 was classified as a cataclysmic variable \citep{Torres2014}, but with a radio counterpart in this survey it is possible that it is a quiescent black hole X-ray binary. CX330 was classified as a type of young stellar object \citep{Britt2016} based on infrared outburst and variability, with an inverted spectral index radio counterpart. CX51 was not previously classified, but with a very steep spectrum radio counterpart it may involve a pulsar. These three sources are discussed further in Section~\ref{Notable Sources}. The remaining 16 sources are currently unclassified, although CX480's radio counterpart with a very steep spectral index is notable. These X-ray matched sources may be revisited to determine if they are Galactic sources, and to attempt further classification. 

We also note that \citet{Maccarone2012} and \citet{Jonker2014} previously matched CXOGBS sources to NVSS, six of which are in the region covered by our radio survey and are all detected, as noted in the Table~\ref{table:CXOmatches}. All six of these previously matched sources are AGN, with CX488 being an interesting case. The position match and spectral index of CX488 may indicate that the AGN has turned off some time ago, as all of the extended structure is rather steep spectrum and there is no clear location of the nucleus in the radio image that is compatible with the X-ray location. CX480 has a radio counterpart with a very steep spectral index, making it a decent pulsar candidate. The potential optical counterpart to CX480 is a very red star, possibly an M~dwarf. If the association is real, CX480 may be a pulsar in a binary with this M~dwarf, or the association could be a coincidence given the number density of dwarf stars in the Galaxy.

\subsection{{\it Fermi}-LAT match}

We also matched our radio catalog to gamma-ray data from the {\it Fermi}-LAT mission \citep{Atwood2009Fermi}. We identify 16 {\it Fermi} sources that overlap with our survey area. In total, over these 16 {\it Fermi} sources, there are 101 radio sources within the 95\% {\it Fermi} error ellipses. The number of radio sources per {\it Fermi} source ranges from 1 to 25, which is strongly dependent on the location of the {\it Fermi} source relative to the fields, i.e. {\it Fermi} sources near the edge of the fields will not have as many radio sources due to the decreased primary beam response, and may also partially extend beyond the survey region as well. The {\it Fermi} sources are listed in Table~\ref{table:Fermi}, including notes about various radio sources that are within the search radius of the {\it Fermi} source. We do not claim that any particular radio source in this survey is a true association to a given {\it Fermi} source; this match was done mainly because {\it Fermi} is a significantly important catalog in this region of the Galaxy, there are a limited number of known gamma ray sources, and the main gamma ray emitting source classes of pulsars and blazars (beamed AGN, \citealt{Abdo2010}) often have radio counterparts.

\subsection{MIPSGAL match}

Matching our radio catalog to the mid-infrared sky survey MIPSGAL  \citep{Carey2009} yielded 118 sources within a 1" match radius, with the peak at $\sim$0.4", and we estimate a chance match rate of $\sim$1\%. 
The 31 known planetary nebulae in the survey are a significant fraction of mid-infrared and radio matches, and the remaining are likely to be AGN, though additional unknown planetary nebulae and young stellar objects in this region of the Galaxy are possible as well. The MIPSGAL matching also reinforces a small $\sim$0.5" positional error for VLAGBS sources.

\subsection{AGN estimate} \label{subsec:AGN}

While the focus of the radio survey is to identify Galactic sources by choosing a dense region of the Milky Way, a large fraction of sources are expected to be AGN. We matched our radio sources to AGN catalogs such as Million Quasars \citep{Flesch2015}, but this did not result in any matches of known AGN, which we attribute to our location in the Galactic Bulge and its dense population of reddened stars. 
We also estimate how many AGN (or other sources of galaxy radio emission) are expected to be in the survey region by comparing to the data of the VLA 1.4~GHz Survey of the Extended Chandra Deep Field-South \citep{Miller2008,Miller2013} (VLA ECDF-S). These results are presented in Table~\ref{table:AGN_est_Miller}. We scaled the counts of the VLA ECDF-S sources to the larger survey area size of the VLAGBS, and applied an additional scaling factor for the two lowest flux bins due to the effect of the primary beam on the VLAGBS. These results indicate that indeed almost all radio sources in the VLAGBS are expected to be of extragalactic origin, though we reiterate that the extrapolated numbers are estimates from a much smaller survey sky area.



\section{Notable Sources} \label{Notable Sources}

We have identified individual sources of interest based on a variety of factors. Unless stated otherwise, flux densities and angular sizes stated are centred and as they appear in the full bandwidth images at 1.5~GHz, respectively. Images are included for extended sources, with the color scales and contours used to highlight the structure of the sources. The synthesized beam is shown in the lower left corner of each image.
The source images are from the fields without a primary beam correction. The color scales use a log scale, and colorbars are provided at the top of each image, with the flux density values corrected for the primary beam response. The contour level values are reported in the caption of each image in units of mJy, and these values are also primary beam corrected. The contour levels of each source are equivalent to 3, 5, 10, 20, 50, 100, 500, and 1000$\sigma$. Only contours present in the individual image are reported in the caption.

\subsection{CX63 (J173411.3--293117): a radio-emitting compact binary} 
CX63 is a CXOGBS source that was classified as a cataclysmic variable (CV), an accreting white dwarf binary system, based on its accretion-dominated optical spectrum and optical variability \citep{Torres2014}. There is a radio counterpart for this source with a flux density of 0.24$\pm$0.04~mJy, and spectral index of --0.80$\pm$0.90. In optical and X-rays, CVs and quiescent black hole X-ray binaries (qBHXBs) appear to be very similar to each other. Since CVs are far more numerous than BHXBs, and because the optical spectrum displays narrow emission lines, the most likely classification was as a CV. While 
some CVs are known to emit radio from a synchrotron jet \citep{Coppejans2020}, or via electron cyclotron maser emission \citep{Barrett17},
few CVs reach the radio/X-ray flux ratios of qBHXBs \citep{Ridder2023}. 
The radio counterpart is bright compared to the X-ray source, slightly above the radio-bright BHXB track; assuming a distance of 1kpc, the 
1-10~keV intrinsic X-ray luminosity is $1.6 \times 10^{31}$ erg s$^{-1}$, and the 5~GHz (assuming a flat spectrum) radio luminosity is $1.4 \times 10^{27}$ erg s$^{-1}$. 
If this source is located at or more than $\sim$1kpc away, the brightness of the radio emission would be among the brightest of any CV to date. 
Crucially, the radio/X-ray flux ratio appears to be higher than for any known CV, although the X-ray and radio data were taken 4 years apart.

Arguing against a quiescent BHXB nature is the narrow width of the emission lines as presented in \citet{Torres2014}, where H$\alpha$ has a full-width at half-maximum (FWHM) of only 14.8~\AA, or 677~km/s. Comparing this with the quiescent BHXB and CV H$\alpha$ emission lines catalogued in \citet{Casares2015}, we see that the 12 quiescent BHXBs have FWHM of 1000-2850 km/s, while the 28 CVs have FWHM of 341-2460 km/s. The H$\alpha$ line thus provides mildly suggestive evidence in favor of the hypothesis that CX63 is a CV.  A BHXB observed at a very low inclination angle could also have similarly narrow lines as well and explains the radio emission more naturally.


If this is a quiescent BHXB, this source would be even more interesting in that it will have been discovered while in quiescence, as it would be the first confirmed BHXB to be discovered in this way.
A quasi-simultaneous X-ray/radio measurement could confirm this amazingly high radio/X-ray ratio. Estimation of the compact object's mass function from a radial velocity curve, combined with inclination angle estimates, 
would conclusively determine 
whether the accretor is a WD, NS, or a BH. 

\subsection{CX330 (J173643.8--282121): a young stellar object-type source} 
CX330 is a CXOGBS X-ray source that was classified as a possible type of young stellar object from multiwavelength data, though it is not located near a known star-forming region of the Galaxy. It was observed to host a persistent dusty envelope while undergoing a several-years-long infrared outburst that likely began in 2009 \citep{Britt2016}. Optical spectra taken in 2016 showed a continued significant infrared excess. The radio counterpart for this X-ray and infrared source has a flux density of 0.69$\pm$0.07~mJy and a spectral index of +0.86$\pm$0.38, which are typical of ``thermal'' jets seen in young stellar objects \citep{Anglada2018}. However, there is no source at this location in RACS or VLASS (3~$\sigma$ of 1.59mJy; 0.45mJy). The optical/infrared emission has continued to fade since the outburst, thus it is reasonable that the radio emission also faded in the 3 years between VLAGBS and VLASS 1.1, resulting in a VLASS non-detection, especially if the spectral index has also steepened due to cooling/aging of a synchrotron-emitting electron population \citep[e.g.,][]{Ferrari2008,Feretti2012}.

\begin{table*}
    \setlength{\tabcolsep}{10pt}
	\begin{tabular}{cccccl}
	    \hline
	    VLAGBS source & S$_{\mathrm{1.5GHz}}$ (mJy/beam) & $\alpha$ & RACS$_{3\sigma}$ & VLASS$_{3\sigma}$ & Notes  \\
		\hline\hline
		J173454.9--293245 & 0.97$\pm$0.09 & --1.10$\pm$0.38 &  $<1.70$ & $<0.41$ & \\ 
		J173630.7--285210 & 1.25$\pm$0.10 & --0.85$\pm$0.33 &  $<1.52$ & $<0.39$ & \\ 
        J173912.1--272650 & 4.62$\pm$0.26 & --1.10$\pm$0.49 &  $<1.03$ & $<0.35$ & \\ 
		J174029.9--264305 & 1.00$\pm$0.12 & --0.77$\pm$0.99 & $<1.67$ & $<0.36$ & *Nearby Fermi source, tMSP candidate? \\ 
		J174121.8--270721 & 0.95$\pm$0.08 & 0.64$\pm$0.32 & $<1.18$ & $<0.36$ & \\ 
		\hline						
	\end{tabular}
	\caption{A sample of sources that are visible in our survey but not in VLASS and RACS, as described in Section~\ref{Radio Transients}. All sources are point sources. Flux densities are in mJy. The $3\sigma$ upper limits of RACS and VLASS are provided (and none are visually apparent in any of the three VLASS epoch images). For the source marked with an *asterisk, see Section~\ref{Notable Sources} for a further discussion.}
	\label{table:GBStransients}
\end{table*}

\begin{table*}
    \setlength{\tabcolsep}{10pt}
	\begin{tabular}{cccccl}
	    \hline
	    VLAGBS source & S$_{\mathrm{1.5GHz}}$ & $\alpha$ & VLASS 1.1 & VLASS 2.1 & VLASS 3.1\\
		\hline\hline
		J173155.5--302304 & 0.94$\pm$0.07 & -0.42$\pm$0.28 & 0.47$\pm$0.13 & 0.89$\pm$0.14 & 0.74$\pm$0.12\\ 
        J174110.2--271300 & 1.02$\pm$0.08 & 0.44$\pm$0.44 & 0.53$\pm$0.12 & 0.71$\pm$0.14 & 0.45$\pm$0.13 \\ 
		J174248.1--264958 & 1.06$\pm$0.08 & -0.44$\pm$0.32 & 0.44$\pm$0.12 & 0.81$\pm$0.13 & 0.62$\pm$0.13\\ 
		\hline						
	\end{tabular}
	\caption{Three sources that appear to be variable over the three epochs of VLASS images. These sources were originally chosen as potential transients (i.e. Table~\ref{table:GBStransients}) due to a non-detection in VLASS epoch 1.1 catalog. Flux density values are from analysis of images from the Cirada image cutout server. Flux densities given are peak values in mJy, and errors are the local RMS value.}
	\label{table:VLASSvariables}
\end{table*}

\begin{table*}
    \setlength{\tabcolsep}{10pt}
	\begin{tabular}{cc|ccc}
	    \hline
	    RACS & & VLAGBS   \\
	    Source & S$_{\mathrm{peak}}$ & 3$\sigma$ & RACS/VLAGBS$_{3\sigma}$ & $\alpha$ lim.  \\
		\hline\hline
		J173218--301236 & 1.504 & $\lesssim$0.08 & 19.39 & $\lesssim$--5.65 \\ 
		J173241--302831 & 1.799 & $\lesssim$0.17 & 10.68 & $\lesssim$--4.51 \\ 
		J174431--262415 & 3.236 & $\lesssim$0.30 & 10.78 & $\lesssim$--4.53 \\ 
		J174627--254425 & 3.225 & $\lesssim$0.29 & 11.12 & $\lesssim$--4.58 \\ 
		\hline						
	\end{tabular}
	\caption{RACS sources that are not found in our survey. Because our survey 
 has much better 
 sensitivity, these sources are either highly variable, or have very steep spectral indices (or a combination). RACS sources that were larger than our survey's largest angular scale were not included in the search. All of the RACS sources are consistent with a point source in the RACS images, and because of this we choose to use the peak flux rather than the total, which is the metric we found to be more reliable for the flux of relatively faint point sources. All of these RACS sources are also not present in VLASS epochs 1.1 and 2.1. Flux densities are in mJy. See Section~\ref{Radio Transients}.}
	\label{table:RACStransients}
\end{table*}

\subsection{CX51 (J174000.6--274859): steep spectrum with X-ray and MIPSGAL counterparts} 
CX51 is an X-ray source identified in the CXOGBS. It has a radio counterpart in our survey that is a point source with a flux density of 0.97$\pm$0.09~mJy and spectral index of --2.31$\pm$0.54. 
It has a MIPSGAL 24~$\mu$m counterpart of 3.761~mJy/8.22 mag at a match radius of 0.47", and an IRAC 4.5~$\mu$m counterpart of 13.18 mag for the same MIPSGAL source. There have been no optical counterparts detected, and with the optical upper limits (e.g., $\sim$21~mag from {\it Gaia}), 
this spectral shape rules out a heavily reddened star at any distance, indicating there must be a significant mid-infrared excess present. 
Bright infrared emission is common among young stellar objects, but the very steep spectral index is difficult to explain in this scenario. The steep spectrum of the radio source suggests a pulsar, but the mid-infrared counterpart strongly disfavors this scenario.
Finally, all counterparts could originate from an AGN with unusually steep radio emission that is obscured enough to not be detected by {\it Gaia} ($\gtrsim 21$ mag), likely at high redshift.


\subsection{VLAGBS J174029.9--264305: a tentative transitional millisecond pulsar (tMSP) candidate} 
This radio source is 1.00$\pm$0.12~mJy, with a spectral index of $-0.77\pm$0.99. It appears to be transient, as it is not in the RACS or VLASS surveys (3$\sigma$ limit [expected]: RACS 1.44~mJy [1.49~mJy]; VLASS 0.41~mJy [0.58~mJy]). There are no optical/infrared counterparts.
The location is just outside of the 95\% error ellipse of the unassociated 4FGL source J1740.7--2640 (4' distance). We also note that the radio source is only $\sim$0.7' from the centre of the previous 3FGL detection of the same Fermi source, J1740.5--2642. The gamma-ray source has been labeled as a millisecond pulsar candidate based on its gamma-ray spectrum \citep{Bartels2016}, but to date there has not been a radio pulsar found inside or near the error ellipse. The Parkes Multibeam Pulsar Survey at 1.4~GHz covered the Galactic Plane, and this area of the sky was reported in \citet{Lorimer2006} with no detection of this source. Altogether, there is the possibility that this source is in fact a millisecond pulsar, in a tMSP system that was in the pulsar state during VLAGBS and transitioned to an accreting X-ray binary state before VLASS. 

Unfortunately, the X-ray and optical observations of the {\it Chandra} GBS do not cover this source, as it is located at the outer edge of the radio survey region that extends beyond the optical and X-ray GBS survey areas. A 2~ks {\it Swift}-XRT observation in April 2023 did not detect an X-ray source at this location, with a flux upper limit of $2\times10^{-13}$~erg~s$^{-1}$~cm$^{-2}$ or 10~kpc distance luminosity upper limit of $2\times10^{32}$~erg~s$^{-1}$, ruling out an active X-ray binary presence at the time.

\begin{table*}
	\begin{tabular}{cc|ccc|cccc}
	     \hline
	     VLAGBS source & S$_{\mathrm{1.5GHz}}$ & MIPSGAL & 24$\mu$m &  & SPITZER & bands & (mag) & \\
		  &  mJy &  & mJy & mag & 3.6$\mu$m & 4.5$\mu$m & 5.8$\mu$m & 8.0$\mu$m\\
		\hline\hline
		J173526.3--291723 & 1.17$\pm$0.08 & MG358.5208+01.6977 & 80.38 & 4.89 & 11.08 & -- & 11.23 & --  \\ 
		J173245.5--300111 & 0.84$\pm$0.07 & MG357.5913+01.7909 & 18.66 & 6.48 & 13.15 & 12.80 & 11.82 & 9.77 \\ 
		J174100.9--265329 & 0.52$\pm$0.05 & MG001.2063+01.9392 & 11.57 & 7.00 & 13.21 & 12.54 & -- & 10.98 \\ 
		J174030.6--270927 & 0.95$\pm$0.08 & MG000.9207+01.8940 & 17.28 & 6.56 & -- & -- & 12.48 & 10.23 \\ 
		\hline						
	\end{tabular}
	\caption{Potential new planetary nebulae, with the MIPSGAL and supplied Spitzer/IRAC infrared measurements from the matched source. These sources are extended and rounded in the radio images and have a comparatively bright MIPSGAL counterpart. Images are in Image~\ref{imagePNe}.}
	\label{tab:tablePNe}
\end{table*}

\begin{figure*}
    \begin{tabular}{cc}
    \captionsetup[subfloat]{justification=centering}
    \subfloat[J173454.9--293245 \\Contours (mJy): 0.10, 0.17; peak 0.36~mJy; total 0.97~mJy ]{\includegraphics[width=3.2in]{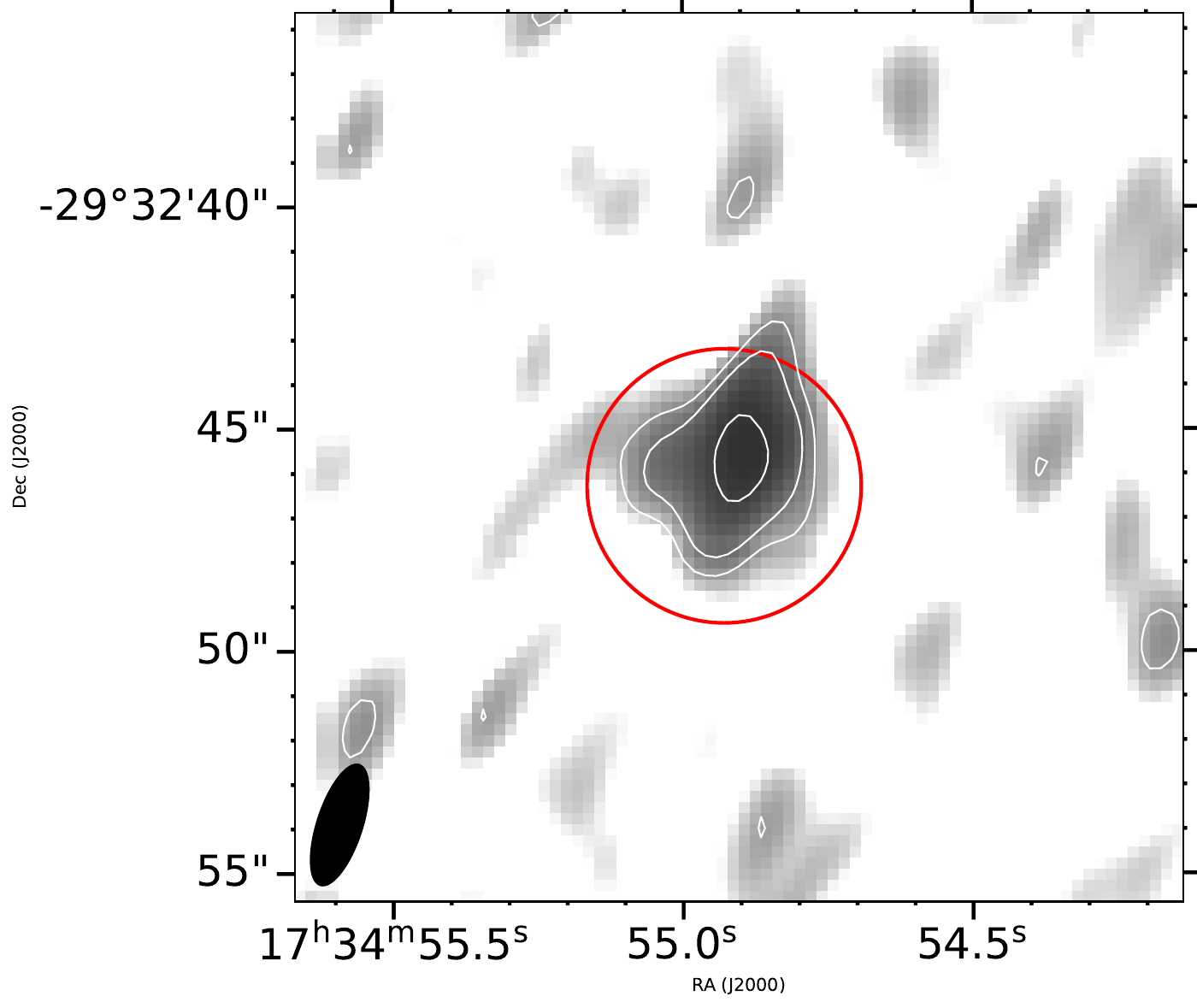}} & 
    \captionsetup[subfloat]{justification=centering}
    \subfloat[J173526.3--291723 \\Contours (mJy): 0.07, 0.11, 0.23; peak 0.31~mJy; total 1.17~mJy ]{\includegraphics[width=3.2in]{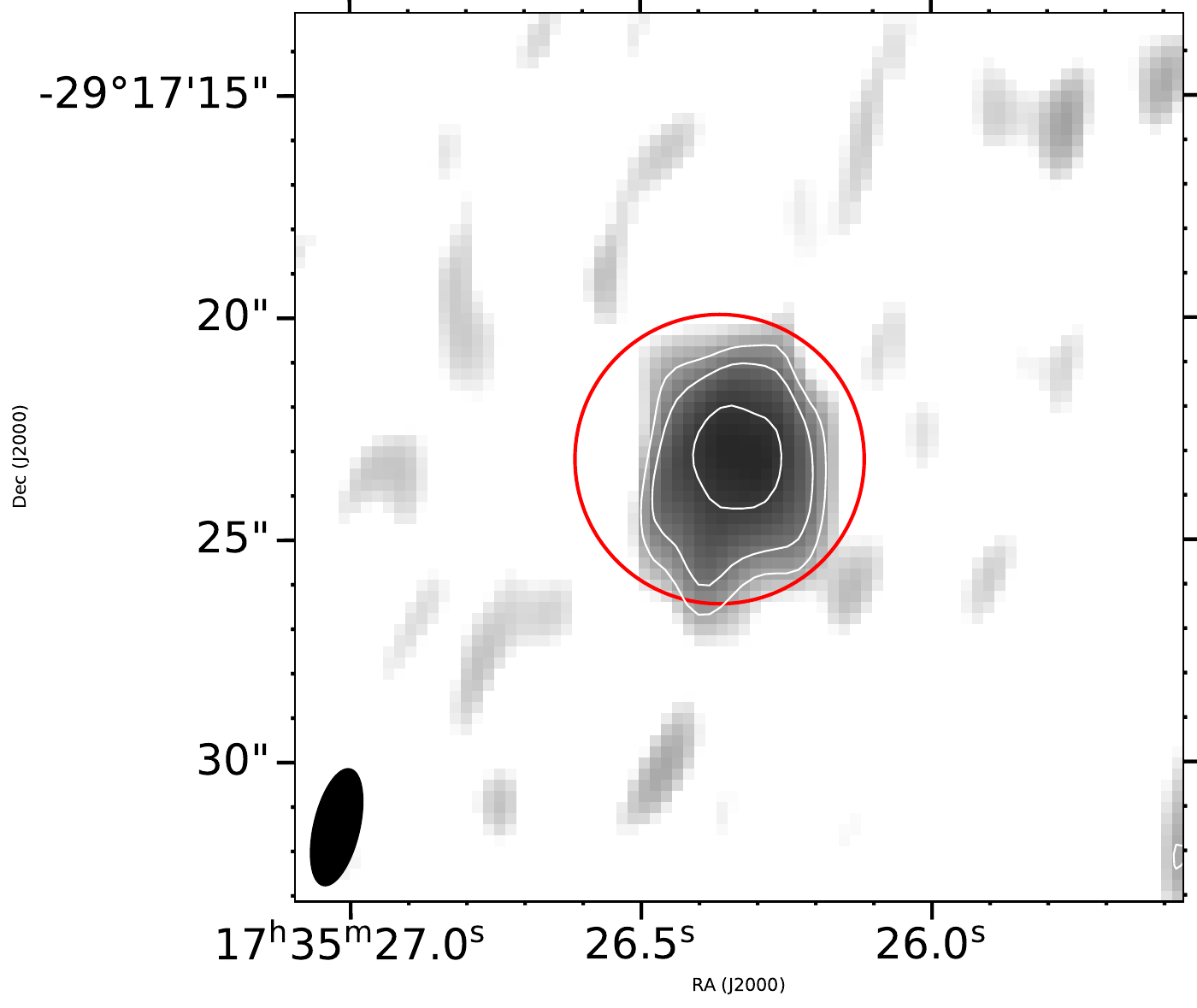}} \\ \\
    \captionsetup[subfloat]{justification=centering}
    \subfloat[J174030.6--270927 \\Contours (mJy): 0.08, 0.13, 0.25; peak 0.46~mJy; total 0.95~mJy ]{\includegraphics[width=3.2in]{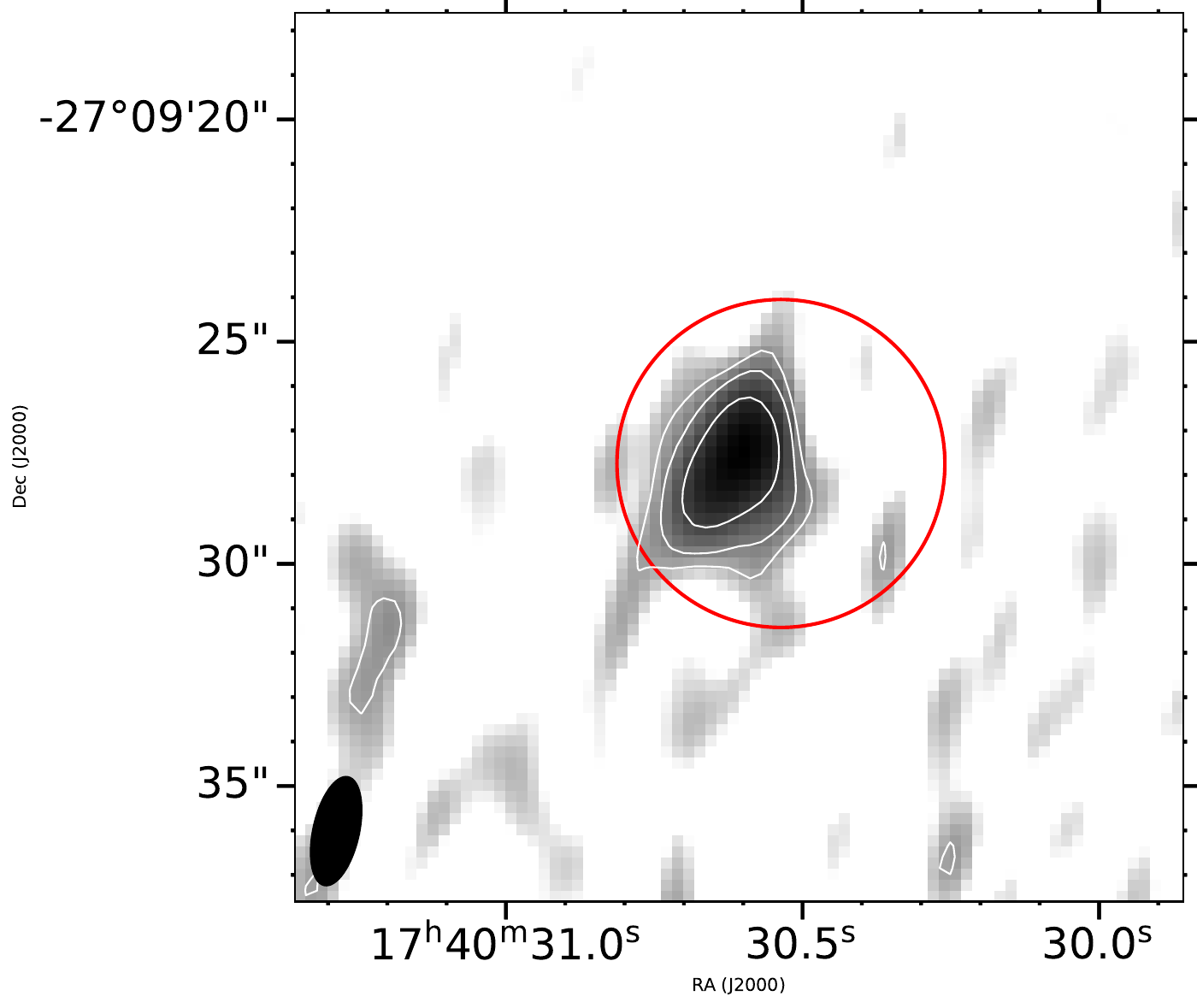}} &
    \captionsetup[subfloat]{justification=centering}
    \subfloat[J174100.9--265329 \\Contours (mJy): 0.07, 0.11, 0.25; peak 0.31~mJy; total 0.52~mJy ]{\includegraphics[width=3.2in]{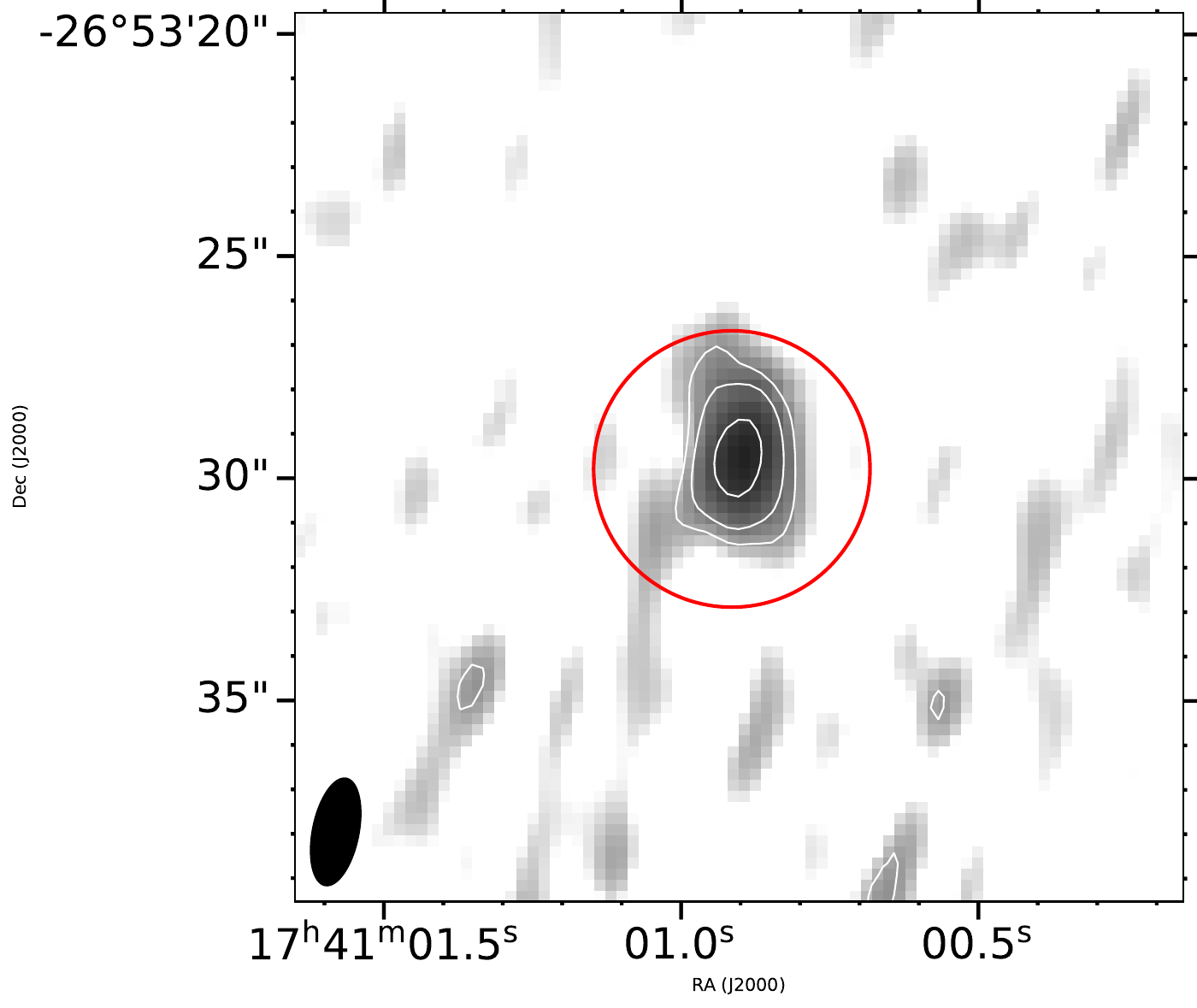}} \\ 
    \end{tabular}
    \caption{Potential new/uncatalogued planetary nebula candidates. Raster is of radio data, and red circles indicating the location and full width half max size of the matched MIPSGAL source. These radio sources are rounded in appearance, and are matched to a relatively bright MIPSGAL 24$\mu$m counterpart, with the ratio of radio to infrared flux following the SED of planetary nebulae, aligning with the known planetary nebulae found in this survey. See Section~\ref{potentialPNe} and Table~\ref{tab:tablePNe} for more information about these sources. For all images, the contour levels of each source are equivalent to 3, 5, 10, 20, 50, 100, 500, and 1000$\sigma$. Only contours present in the individual image are reported in the caption.}
    \label{imagePNe}
\end{figure*}

\subsection{Transient or variable sources} \label{Radio Transients}
There are sources that appear to be transient or variable in the radio, or have very steep spectral indices, by comparing our survey with RACS-low (year 2020) and VLASS (year 2018/2020/2023 for epoch 1.1/2.1/3.1, respectively). In total, there are 429 RACS sources  that are in the VLAGBS region, and 198  VLASS sources. VLAGBS sources that were expected to be bright enough to be detected in RACS and/or VLASS, but were not, are listed in Table~\ref{table:GBStransients} and Table~\ref{table:VLASSvariables}. RACS sources not detected in VLAGBS are listed in Table~\ref{table:RACStransients}; all VLASS 1.1 sources are recovered.

Sources in RACS but not VLAGBS could be explained by very steep spectra, but this is unlikely, as we do not find any similarly steep spectrum matches between detected VLAGBS and RACS sources: the steepest matched source that is an isolated point source in both surveys has a RACS-VLAGBS spectral index of $-3.5$ (and VLAGBS in-band spectral index of also $-3.5$), and the next-steepest source is $-1.4$ between RACS and VLAGBS. A possibility for any of these point sources is scintillation, which could affect the flux and spectral index enough to cause a nondetection at other frequencies \citep{Rickett1990}. Flare stars could also explain  transient radio emission \citep{Melrose1982,Berger2002}, and the progenitor M~dwarfs could very reasonably be undetected in infrared surveys, considering their faintness and crowding issues when looking toward the Galactic Bulge. With the possibility of flaring stars, we looked at these selected VLAGBS transients in each of the 4 scans per field to determine if there are signs of significant variability over timescales of a few minutes, but none of the sources show strong variability (J173630.7--285210 shows the strongest potential variability, but over the 4 scans its flux amplitude is only 20\%), so flaring stars are an unlikely explanation for  these apparent transients. 


\subsection{Possible planetary nebulae (PNe)} \label{potentialPNe}
There are four extended faint radio sources that have coincident MIPSGAL sources indicating a large 24~$\mu$m flux compared to radio, for a ratio that aligns with known PNe found in the survey region of $\geq$20 (empirically in our survey, brighter PNe tend to have a larger ratio), and that follow typical PNe spectral energy distributions that peak at $\sim$24~$\mu$m \citep{Zhang1991}. These may be previously uncatalogued planetary nebulae (PNe). The ratio of high 24~$\mu$m to 1.5~GHz flux density indicates that these sources are not related to the extended synchrotron lobes or hotspots of AGN (which have infrared/radio ratios of <0.1, e.g., \citep{Heckman1992}), the largest population of extended radio sources for this survey. These sources are fainter than any cataloged radio PNe that are in the survey region as well, and individually are fainter than most or all known PNe at 24~$\mu$m. If any of these are in fact PNe, this indicates that radio observations are suitable for finding fainter PNe candidates throughout the Galaxy, especially as radio observations do not suffer from the same crowding and extinction issues that infrared and optical do. These selected sources are reported in Table~\ref{tab:tablePNe}, and images of the radio sources in Figure~\ref{imagePNe}.

H~II regions do appear similar in radio to PNe. The infrared counterparts to the selected sources are very faint and do not have detections in all of the necessary bands to distinguish  between PNe and H~II regions as in \citet{Anderson2012}, and as such they may be H~II regions rather than PNe. However, PNe are about an order of magnitude brighter at 24~$\mu$m relative to their peak emission, and ultra- and hypercompact H~II regions are often rather inverted in spectral index ($\alpha \simeq 1$ \citep{Beuther2007}), and are thus usually much fainter than PNe at frequencies below 10~GHz. The angular size of the chosen sources here (3'' to 5'') aligns well with the smaller end of the planetary nebulae found in the Galactic Bulge in \citet{Jacoby2004}. There are no sources classified solely as H~II regions in the VLAGBS region; most similar sources are classified as PNe only, and a few are  classified in both PNe and H~II catalogs.

\begin{figure*}
\begin{tabular}{ccc}
\subfloat[J174411.7--255207 \\Total 70.78~mJy; peak: 3.54~mJy ]{\includegraphics[width=3.2in]{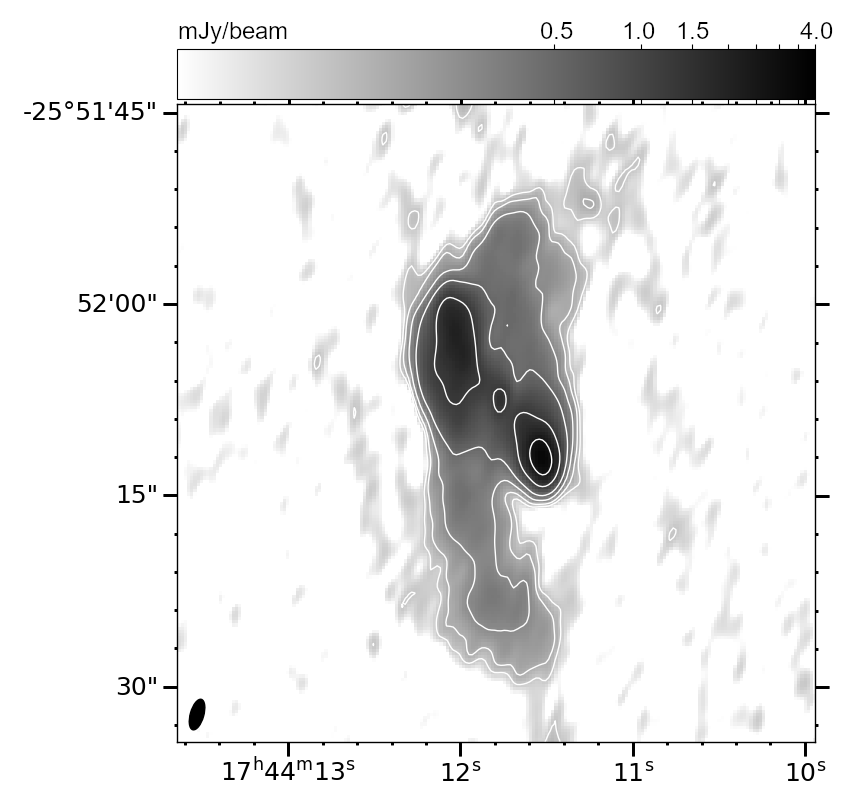}} &
\subfloat[J173635.9--290008 \\Total 134.91~mJy; peak 41.88~mJy ]{\includegraphics[width=3.2in]{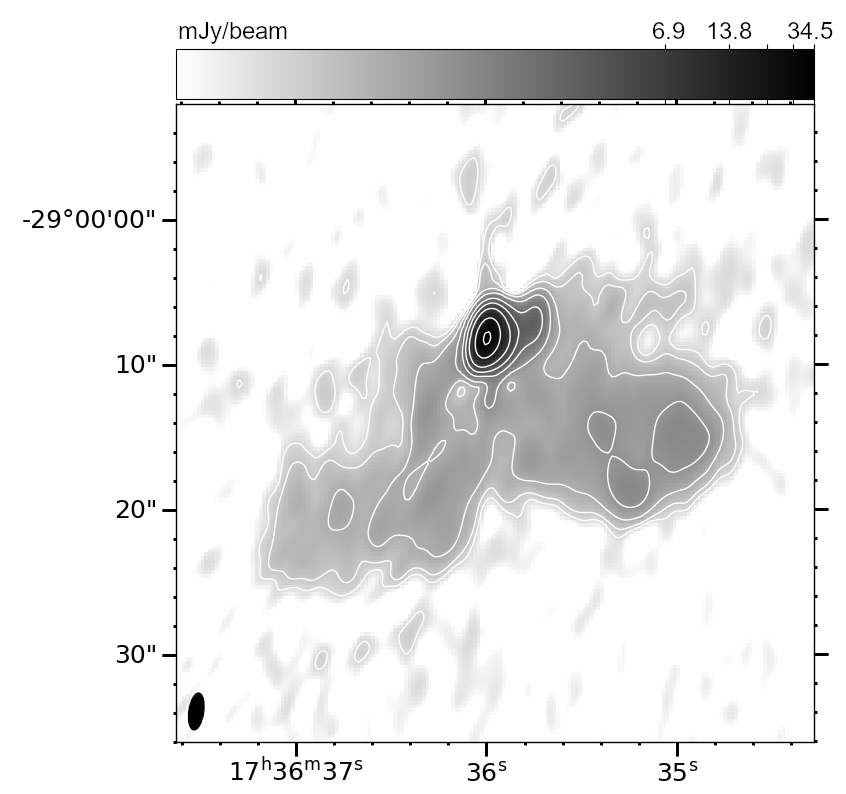}} &
\end{tabular}
\caption{Possible X-shaped galaxies, displaying regions of active jet emission and old jet emission. X-shaped galaxies are believed to be evidence of supermassive black hole mergers. See Section~\ref{XSG}. }
\label{image:XSGs}
\end{figure*}

\begin{figure}
    \centering
    \includegraphics[width=0.45\textwidth]{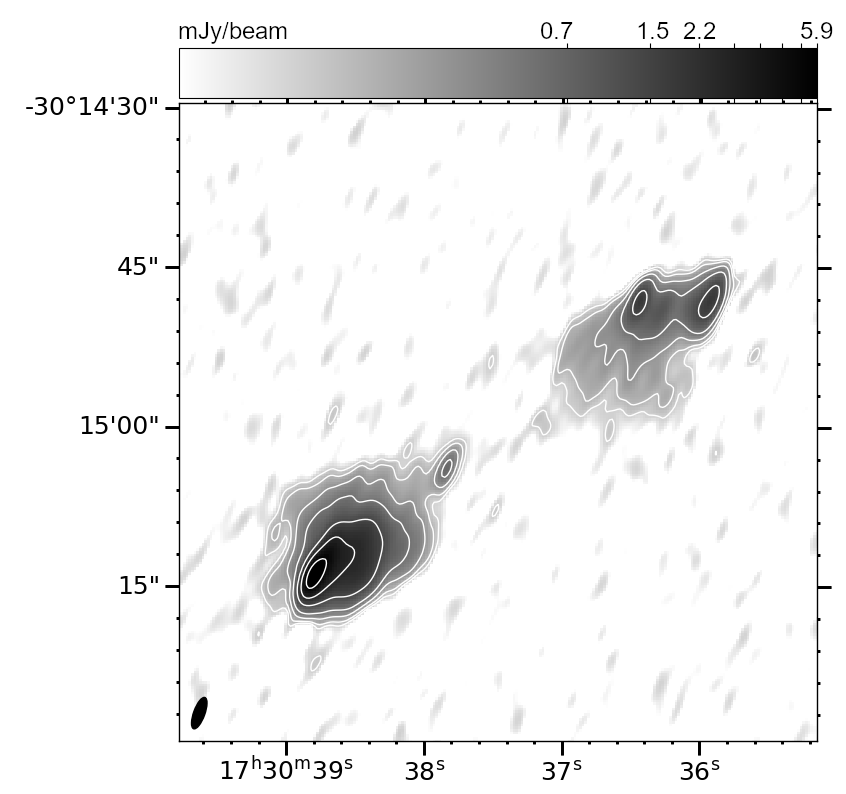}
    \caption{Possibly a single AGN with multiple hotspots, similar to Cygnus~A. See Section~\ref{dualAGN}. \\ Upper/western: total 21.83~mJy; peak 2.34~mJy. Lower/eastern: total 61.68~mJy; peak 9.69~mJy. }
    \label{image:dualAGN}
\end{figure}

\begin{figure}
    \centering
    \includegraphics[width=0.45\textwidth]{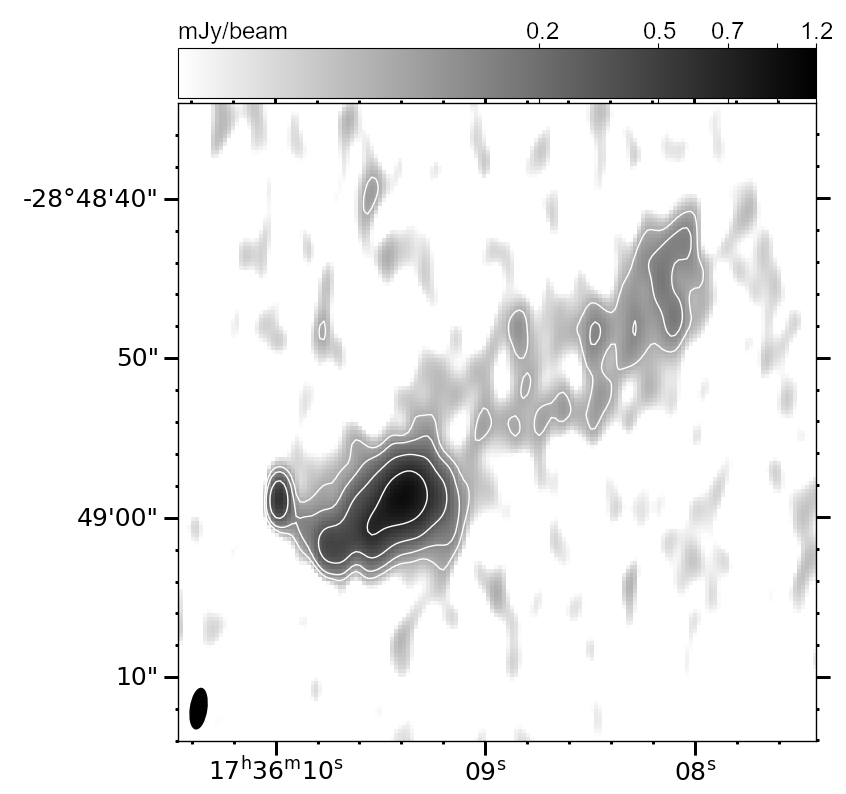}
    \caption{J173609.7--284901: A potential head-tail  galaxy, but also with extended, fainter emission (to the upper right/north-west)} that may be from a previous jet that had turned off for a significant amount of time. \\Contours (mJy): 0.098, 0.163, 0.326, 0.653; peak 1.19~mJy; total 11.23~mJy. See Section~\ref{headTail}.
    \label{image:Head-tail}
\end{figure}

\subsection{X-shaped galaxies (XSGs)} \label{XSG} 
XSGs are a class of AGN that are believed to be caused by the merger of two supermassive black holes (SMBHs), for which the resulting SMBH has a new spin axis orientation \citep{MerrittEkers2002}. The effect of these mergers is seen in radio images as two sets of radio lobes. One set is from the previous emission of a former individual SMBH AGN and appears diffuse and inactive. The other set appears as brighter emission from the resulting SMBH that is active \citep{Cheung2007}. The prototype XSG is NGC326 \citep{Ekers1978}. We find two AGN that may fall into the XSG class (Figure~\ref{image:XSGs}). 
The better candidate, J174411.7--255207, has two clear sets of extended lobes, one of which is fainter and set at an angle to the other brighter set, with a brighter central flux source that is likely the location of the nucleus. The other, less clear candidate, J173635.9--290008, has a ``head'' component, and a flared bidirectional ``tail'' component. The head is made of two point sources (clearly seen in residuals while cleaning the image), one brighter than the other, and the tail is quite large and appears to be swept away from the head. In an XSG interpretation, this galaxy may have been moving though a medium as the merger occurred. 


\subsection{J173036.1--301448 and J173037.9--301505: Cygnus A-like AGN} \label{dualAGN} 
These two extended radio sources may be produced by the same AGN, where the two jets each have two distinct hotspots (Figure~\ref{image:dualAGN}), similar to the radio jets of Cygnus A \citep{Perley1984}. The hotspots and extended emission are not resolved in RACS due to its lower angular resolution (the total source appears as two elliptical flux islands), nor in VLASS due to being at a higher frequency and lower sensitivity (the diffuse emission is not detected; the future image stack of all VLASS epochs will provide better sensitivity). It is also possible that the point source close to the centre of the image is an interloper and unrelated, though there is a connecting flux bridge to the bright hotspot to the southeast that suggests a common emission origin. 

\subsection{J173609.7--284901: A Head-Tail galaxy?} \label{headTail} 
This source (image~\ref{image:Head-tail}) may be a type of head-tail galaxy \citep{Miley1972}. These are AGN that have bright ``heads'' of emission near their nucleus, and additional ``tail'' components that are extended and appear to be swept back, generally leading to the appearance of motion of the AGN. For this source, though, it is unclear where the nucleus is located. It displays two components of extended emission, one that is brighter and appears to be active, and another that is an extension of the brighter lobe-like component, and is fainter and very diffuse in appearance. It may be that the jet of this AGN had been turned off for a significant amount of time, allowing the older emission to fade, and has since reignited its jets, leading to the brighter lobe emission seen. 



\section{Conclusions}\label{Conclusions}

We have reported the results of a $\sim$3 square degree survey in the Galactic Bulge at 1--2~GHz, with high resolution and sensitivity. This effort has returned 1617 unique sources, and each source has a spectral index calculated if possible. We find a population of $\sim$100 point sources that have a steep spectral index, consistent with that of pulsars. While not all of these sources are expected to be pulsars, it is possible that many of these are. As there are only five known pulsars in the survey region, this may be evidence that there are many pulsars throughout the Galactic Bulge and Centre in general that occupy a region of parameter space that timing-based pulsar searches have generally not been sensitive to, such as millisecond pulsars with high dispersion measures. As such, imaging may currently be the best method to investigate the full population of pulsars in the Galaxy. By matching to other catalogs, we find a few particularly interesting sources: a quiescent black hole X-ray binary candidate that was previously classified as a cataclysmic variable; a years-long infrared transient that is also an X-ray and radio source; a possible transitional millisecond pulsar candidate; and a very steep spectrum radio source with X-ray and infrared counterparts. We identify several potential transient radio sources by matching our catalog with RACS and VLASS, and find four sources that may be currently unknown planetary nebulae based on their faint extended radio morphology and coincident bright 24$\mu$m MIPSGAL source. There are also several visually interesting known AGN sources or AGN candidates.

This survey has covered only one portion ($\sim 1/4$) of the {\it Chandra} Galactic Bulge Survey region, which in total viewed about 5\% of the Galaxy's stellar mass. A similar survey to this may be performed several more times in order to cover the totality of the GBS region, and reveal even more interesting sources emitting in radio.

\section*{Acknowledgements}


We thank the anonymous referee for reading the manuscript thoroughly and providing helpful feedback. We also thank Amaris McCarver and Rudy Montez for discussions about separating out planetary nebulae from compact HII regions; Amruta Jaodand for discussion about transitional millisecond pulsars; Pallavi Patil, Anna Kapi\'nska and Amy Kimball for discussion about AGN; Tracy Clarke for assistance with VLITE data analysis; and Alex Tetarenko for useful discussions about both data analysis and source properties. MAPT acknowledge support from the Ministerio de Ciencia e Innovaci\'on under grant PID2021-124879NB-I00.
This research has made use of the CIRADA cutout service at URL cutouts.cirada.ca, operated by the Canadian Initiative for Radio Astronomy Data Analysis (CIRADA). CIRADA is funded by a grant from the Canada Foundation for Innovation 2017 Innovation Fund (Project 35999), as well as by the Provinces of Ontario, British Columbia, Alberta, Manitoba and Quebec, in collaboration with the National Research Council of Canada, the US National Radio Astronomy Observatory and Australia’s Commonwealth Scientific and Industrial Research Organisation.
This research has made use of the VizieR catalogue access tool, CDS, Strasbourg, France (DOI : 10.26093/cds/vizier). The original description of the VizieR service was published in 2000, A\&AS 143, 23.
This research has made use of "Aladin sky atlas" developed at CDS, Strasbourg Observatory, France: 2000A\&AS..143...33B (Aladin Desktop), 2014ASPC..485..277B (Aladin Lite v2), and 2022ASPC..532....7B (Aladin Lite v3).
JS acknowledges support from  NSF grant AST-2205550 and the Packard Foundation.
COH acknowledges support from NSERC Discovery Grants RGPIN-2016-04602 and RGPIN-2023-04264.

\section*{Data Availability}

The data are publicly available under Project Code 15A-073 in the NRAO archive. The 5$\sigma$ catalog is available online, as well as appendix material. Further data products are available upon reasonable request to the corresponding author.



\bibliographystyle{mnras}
\bibliography{example} 





\appendix

\section{PyBDSF analysis}

\begin{figure*}
\begin{tabular}{ccc}
\subfloat[J173547.9--285840]{\includegraphics[width=0.3\textwidth]{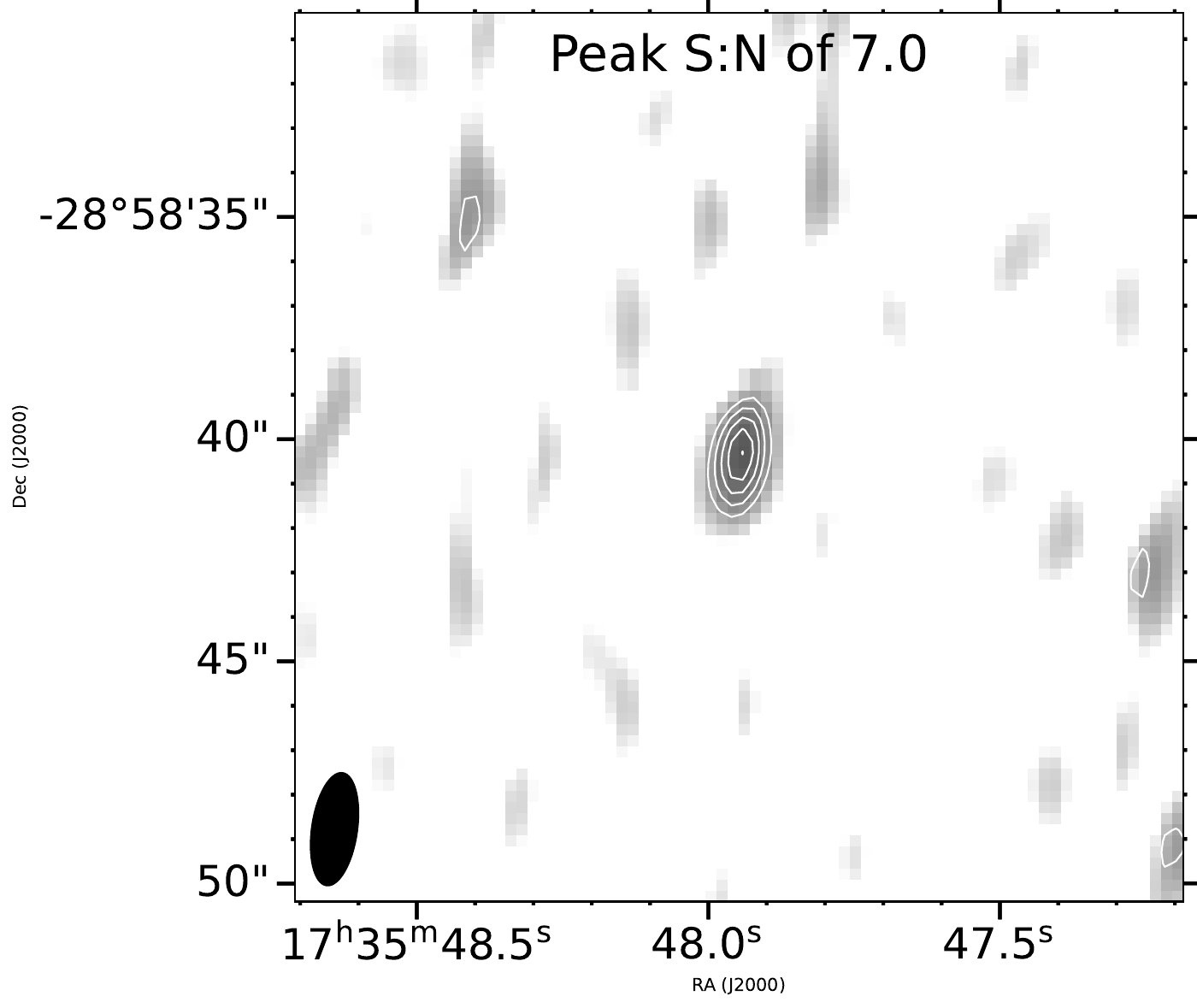}} &
\subfloat[J174115.8--270222]{\includegraphics[width=0.3\textwidth]{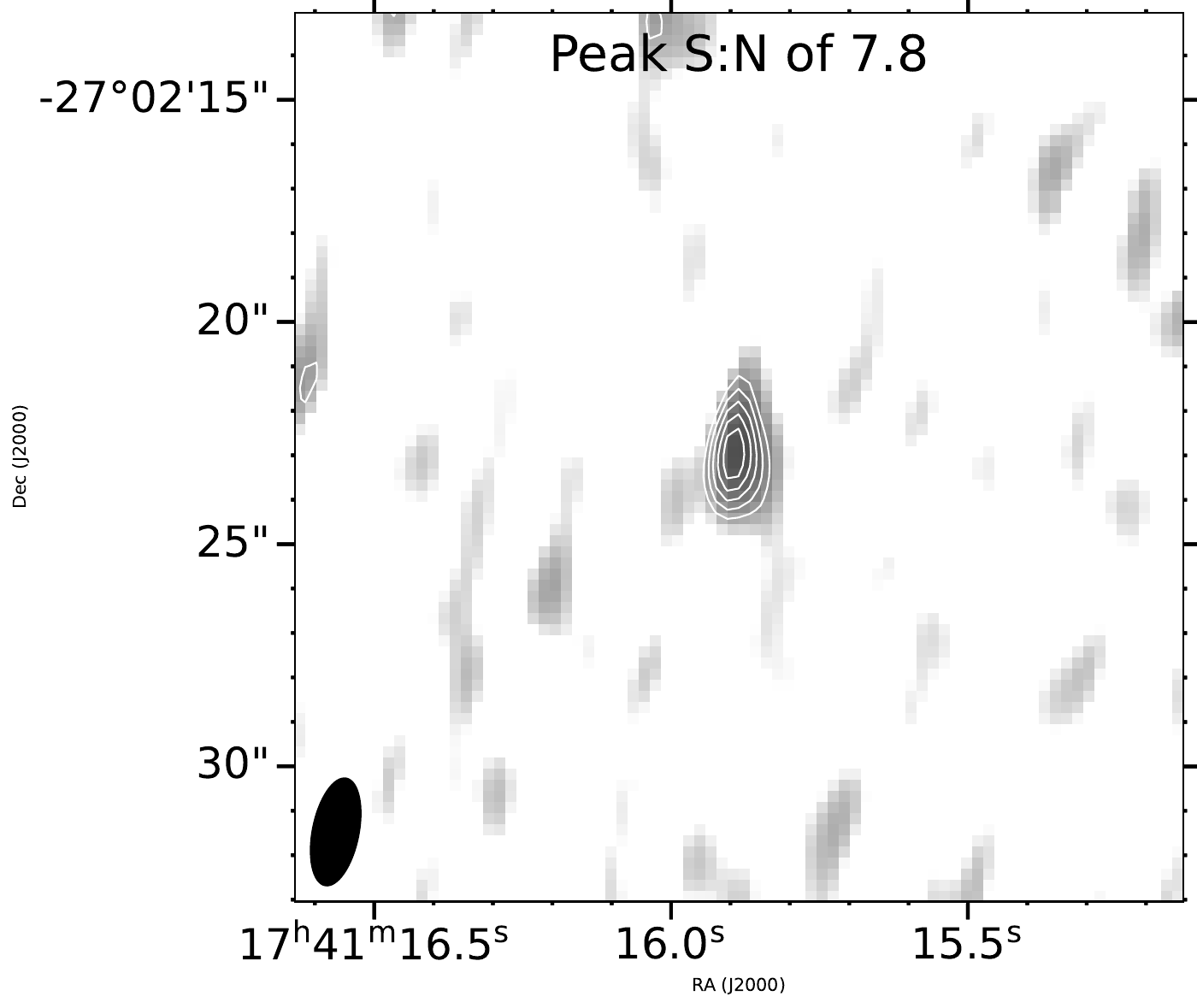}} &
\subfloat[J174530.1--252303]{\includegraphics[width=0.3\textwidth]{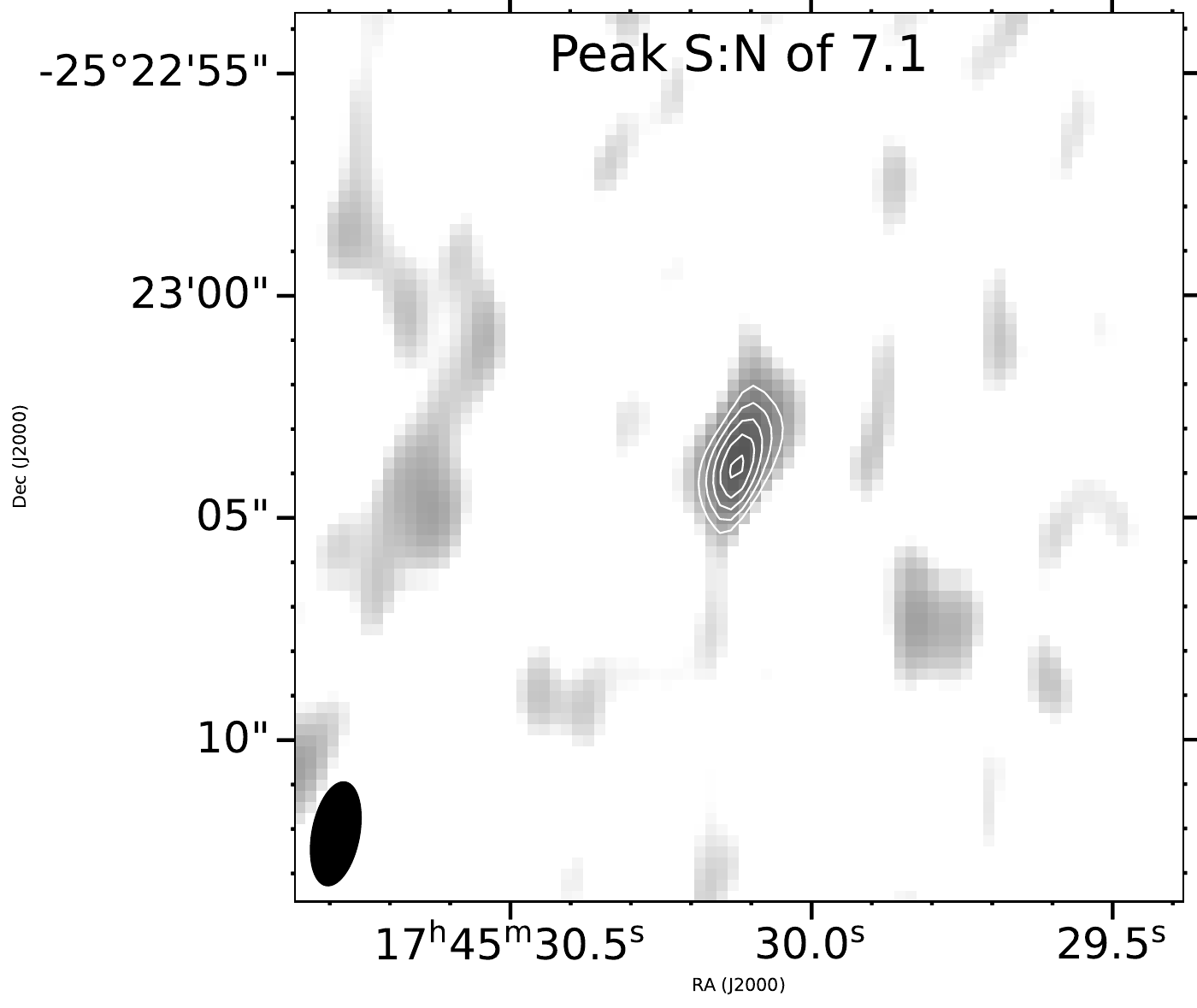}} \\
\end{tabular}
\caption{Sample of statistically significant point sources that PyBDSF did not detect even in the 3$\sigma$ peak and island threshold runs. Contours begin at 3$\sigma$ (3•RMS), increasing by 1$\sigma$ per level.}
\label{image:pybdsf_missed}
\end{figure*}

\begin{figure*}
\begin{tabular}{ccc}
\subfloat[J173157.5--300559 \\Real source on an artifact spoke from a nearby bright source that PyBDSF incorrectly captured, including nearby noise in the total source flux.]{\includegraphics[width=0.3\textwidth]{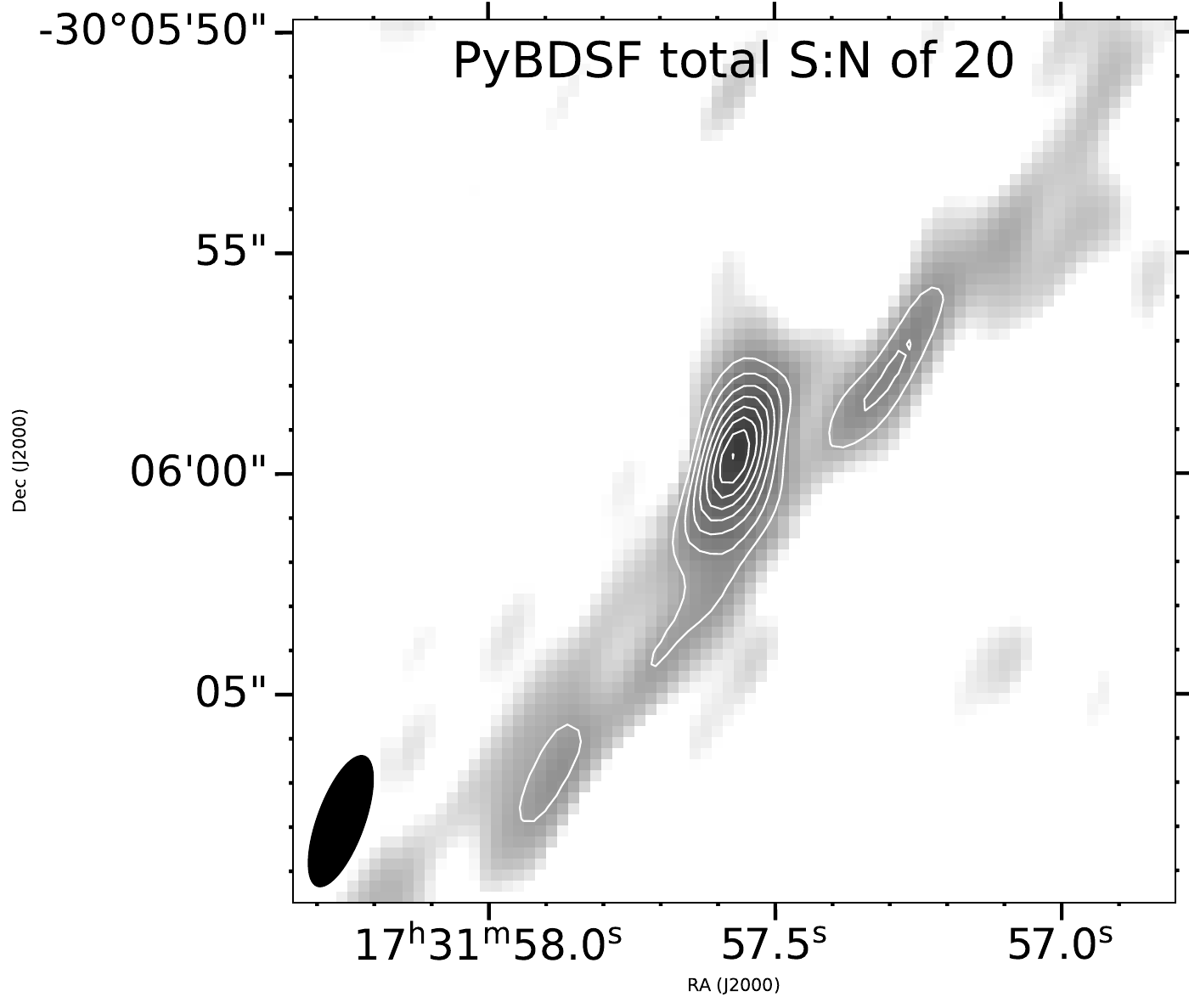}} &
\subfloat[173722.4--280427 \\Extended flux island at the center of image is noise from a nearby bright source.]{\includegraphics[width=0.3\textwidth]{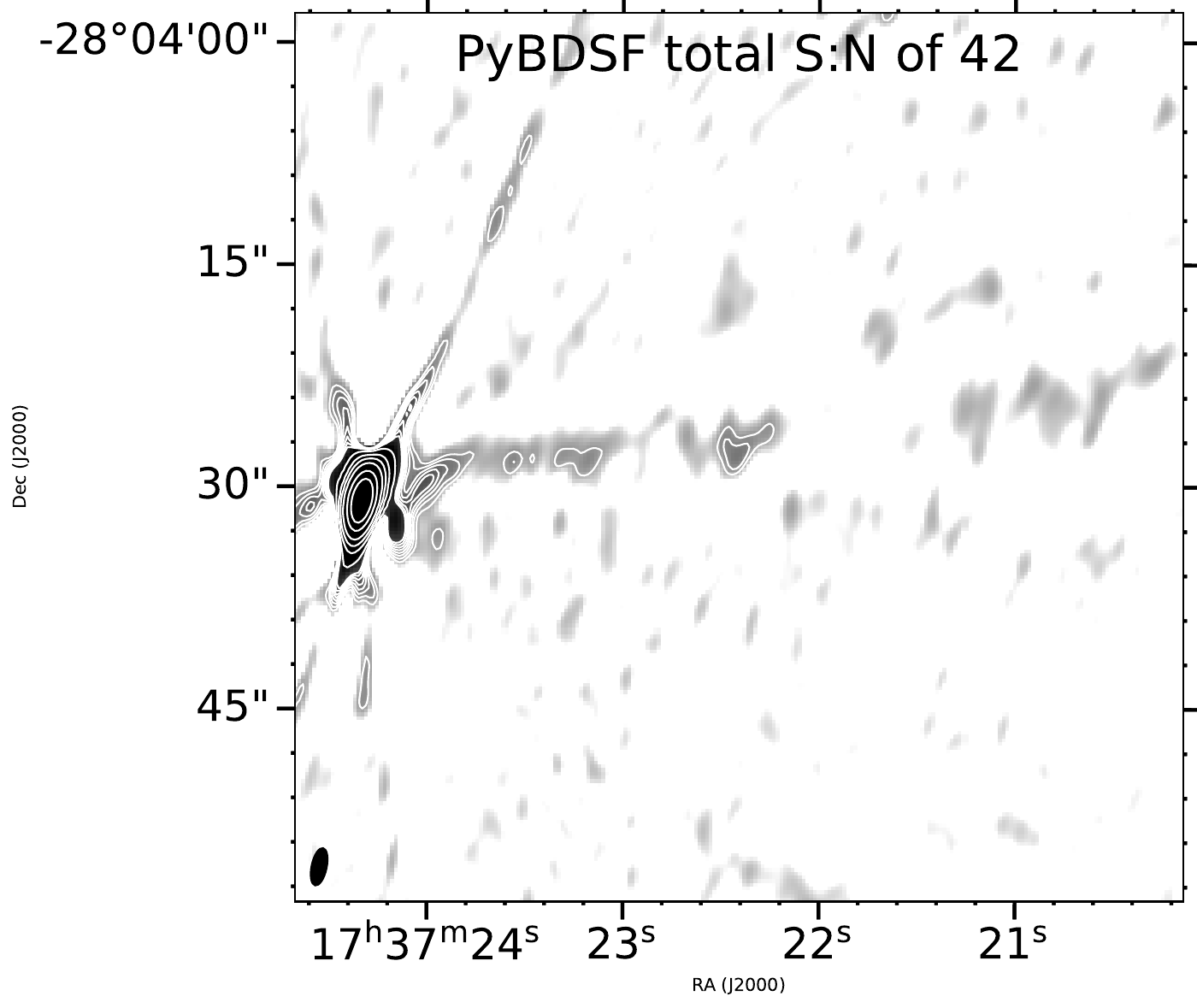}} &
\subfloat[174425.1--255257 \\Line of flux is a radial noise spoke from a bright source off-field.]{\includegraphics[width=0.3\textwidth]{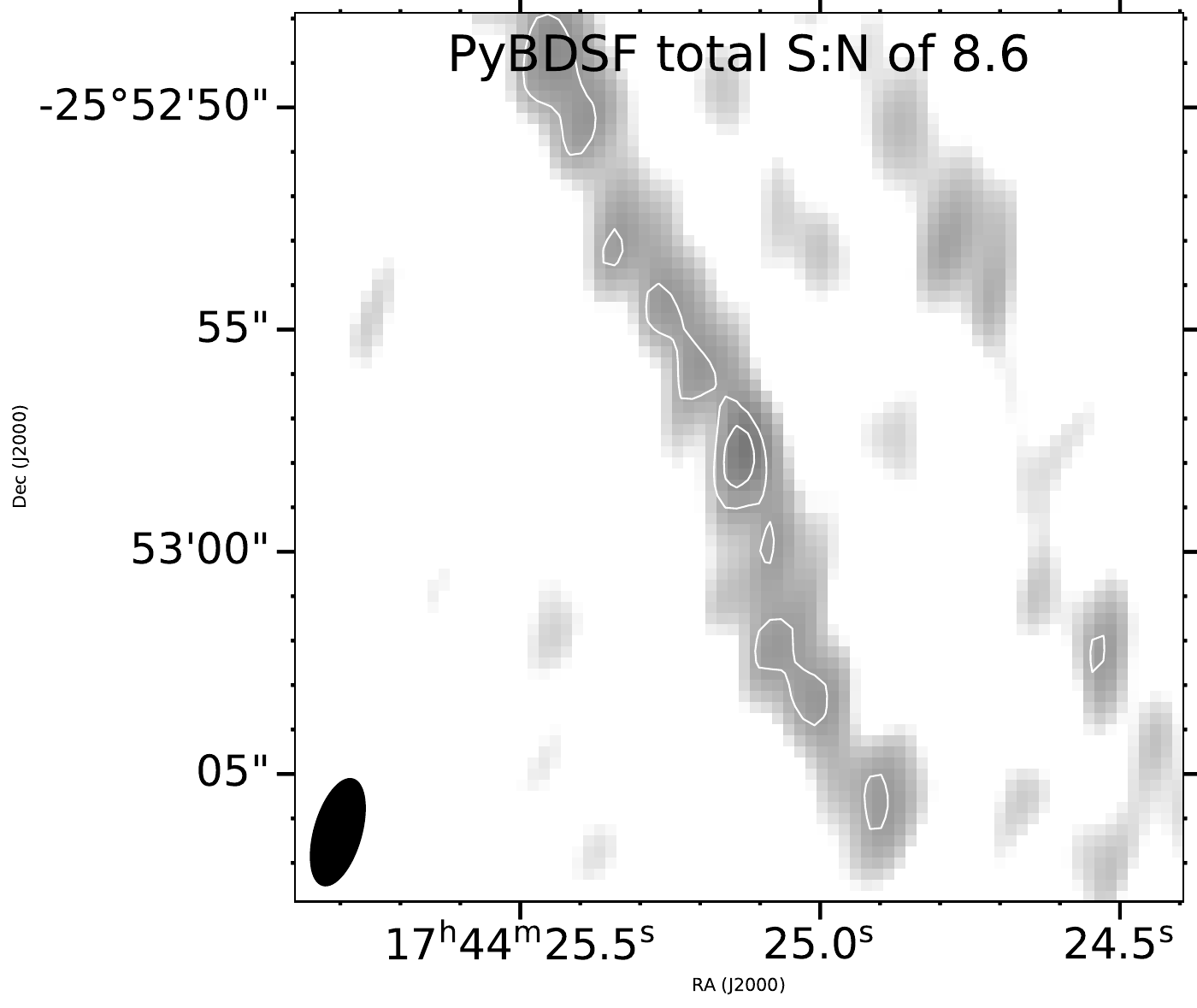}} \\
\end{tabular}
\caption{Sample of sources that PyBDSF identified incorrectly, either because it did not capture the real source correctly, or because the identified flux is noise and not a real source. Contours begin at 3$\sigma$ (3•RMS), increasing by 1$\sigma$ per level until 10$\sigma$, then follow the contour levels described in Section~\ref{Notable Sources}.}
\label{image:pybdsf_errors}
\end{figure*}

We ran PyBDSF several times on the survey fields, each time with \verb+adaptive_rms_box=True+, \verb+rms_box=box+, \verb+rms_map=True+, and \verb+thresh=None+. Five rounds of PyBDSF were run with parameters \verb+thresh_isl+|\verb+thresh_pix+ of 3|3, 4|3, 4|4, 5|4, and 5|5 to investigate what sources it recovered at various thresholds. These results were overlayed on the survey fields and inspected visually. In each 5|5 run there were some sources identified that we strongly believe are noise and not real sources (e.g., on radial artifact spokes emanating from bright sources). In each case also, however, there are a number of statistically significant point sources (peak S:N > 5 calculated manually) that are not detected even in the PyBDSF run of 3|3. It appears that PyBDSF is completely blind to some isolated sources, and we do not know the cause of this blindness. A sample of significant point sources that PyBDSF was unable to recover are shown in Figure~\ref{image:pybdsf_missed}, and a sample of sources that PyBDSF did detect but that we believe to be either inaccurate or noise are shown in Figure~\ref{image:pybdsf_errors}. A smaller number of extended sources were also incorrectly captured by PyBDSF, usually not including large fainter regions next to brighter, more compact flux sources (i.e., it detects the brighter compact source but not the fainter, extended emission attached to the bright component, perhaps calculating the fainter diffuse flux as noise). With these PyBDSF results of sometimes missing statistically significant sources and detecting spurious ones, we continued with the manually assembled VLAGBS catalog.

\begin{table*}
	\centering
	\setlength{\tabcolsep}{10pt}
	\begin{tabular}{cccccc}
		\hline
         & GHz | $\mu$Jy/beam & & & & \\
		Field  & Full & Subband 1 & Subband 2 & Subband 3 & Subband 4 \\
		\hline\hline
	  55 & 1.500 | 19 & 1.122 | 58  & 1.378 | 47  & 1.622 | 58  & 1.878 | 46 \\
        58 & 1.500 | 23 & 1.122 | 64  & 1.378 | 41  & 1.622 | 51  & 1.878 | 39 \\
        61 & 1.500 | 25 & 1.122 | 77  & 1.378 | 40  & 1.622 | 56  & 1.878 | 42 \\
        \arrayrulecolor{lightgray}\hline
        64 & 1.500 | 30 & 1.026 | 136  & 1.346 | 38  & 1.718 | 35  & 1.910 | 44 \\
        67 & 1.500 | 22 & 1.026 | 141  & 1.346 | 32  & 1.718 | 38  & 1.910 | 37 \\
        70 & 1.500 | 23 & 1.026 | 150  & 1.346 | 31  & 1.718 | 41  & 1.910 | 41 \\
        \hline
        73 & 1.500 | 23 & 1.090 | 95  & 1.379 | 48  & 1.591 | 67  & 1.847 | 40 \\
        76 & 1.500 | 30 & 1.090 | 101  & 1.379 | 49  & 1.591 | 86  & 1.847 | 35 \\
        79 & 1.500 | 20 & 1.090 | 89  & 1.379 | 37  & 1.591 | 80  & 1.846 | 35 \\
        \hline
        82 & 1.500 | 20 & 1.122 | 44  & 1.378 | 37  & 1.655 | 49  & 1.878 | 39 \\
        85 & 1.500 | 20 & 1.122 | 55  & 1.378 | 36  & 1.654 | 54  & 1.878 | 35 \\
        88 & 1.500  |  21 & 1.122 | 54  & 1.378 | 33  & 1.654 | 55  & 1.878 | 39 \\
        \hline
        91 & 1.500 | 20 & 1.090 | 79  & 1.378 | 35  & 1.686 | 59  & 1.878 | 51 \\
        94 & 1.500 | 19 & 1.090 | 74  & 1.378 | 36  & 1.686 | 45  & 1.878 | 37 \\
        97 & 1.500 | 22 & 1.090 | 66  & 1.378 | 34  & 1.686 | 51  & 1.878 | 34 \\
        \hline
        100 & 1.500 | 26 & 1.122 | 62 & 1.378 | 58 & 1.622 | 53 & 1.878 | 49 \\ 
        103 & 1.500 | 23 & 1.122 | 70 & 1.378 | 50 & 1.622 | 48 & 1.878 | 36 \\
        106 & 1.500 | 23 & 1.122 | 70  & 1.378 | 35  & 1.622 | 46  & 1.878 | 49 \\
		\arrayrulecolor{black}\hline
	\end{tabular}
	\caption{List of each field of the survey and their four subband images, with central frequency values and central RMS values in $\mu$Jy/beam. Ranges in RMS values are primarily due to spectral window flagging, and in some fields also due to artifacts from bright sources.
	\\ 
    }
	\label{tab:fieldRMS}
\end{table*}

\begin{table*}
	\centering
	\setlength{\tabcolsep}{10pt}
	\begin{tabular}{cc}
		\hline
		5$\sigma$ sensitivity (mJy)  & \% of survey area \\
		\hline\hline
	  0.10 & <1\% \\
        0.15 & 20\% \\
        \arrayrulecolor{lightgray}\hline
        0.20 & 40\% \\
        0.30 & 64\% \\
        0.40 & 80\% \\
        \arrayrulecolor{lightgray}\hline
        0.50 & 92\% \\
        0.60 & 98\% \\
		\arrayrulecolor{black}\hline
	\end{tabular}
	\caption{Estimation of the area of VLAGBS that is sensitive at the 5$\sigma$ level to various flux densities. This was performed by utilizing the central RMS/beam values from Table~\ref{tab:fieldRMS} and applying the effect of the primary beam with distance from the center of the field.
	\\ 
    }
	\label{tab:sensitivity}
\end{table*}

\begin{table*}
    \setlength{\tabcolsep}{5pt}
	\begin{tabular}{ccccccccl}
	    \hline
	    Source & ID & dupl & Total flux & Spectral index & Peak flux & Major & Minor & Notes \\
         &  &  & mJy &  & mJy & '' & '' &  \\
		\hline\hline
		J174351.2--261058 & 2869 & 1 & 220.126$\pm$11.06 & --1.294$\pm$0.13 & 170.947$\pm$0.06 & 3.14 & 1.33 & \\
        J173620.6--283551 & 1112 & 1 & 218.326$\pm$10.94 & --0.952$\pm$0.13 & 105.377$\pm$0.03 & 13.90 & 12.88 & 73A AGN \\
        J174201.8--271309 & 2252 & 0 & 194.154$\pm$9.77 & --0.942$\pm$0.16 & 86.154$\pm$0.06 & 24.35 & 10.88 & 91C - AGN HS \\
        J174000.6--274816 & 1677 & 1 & 174.771$\pm$8.77 & --0.921$\pm$0.14 & 19.182$\pm$0.04 & 23.63 & 9.15 & AGN \\
        J173635.9--290008 & 1037 & 0 & 134.913$\pm$6.78 & --0.374$\pm$0.11 & 41.878$\pm$0.04 & 31.38 & 16.22 & 70D - AGN head + tails? \\
		\hline						
	\end{tabular}
	\caption{A sample of the non-duplicated $\geq5\sigma$ radio catalog (5 brightest in total flux). Source is the J2000 coordinate. ID is the unique ID of the catalog entry. Dupl is the duplicate flag marker: 0 is the only occurrence of the source in the survey, 1 is the primary entry for a duplicated source (closest to the centre of its respective field), 2 is the non-primary entry for a duplicated source. Total flux of the source with errors of the local RMS + 5\% of the total flux. Spectral index of the source with errors from fitting a power law to the subband detections of the source. Peak flux with error from the local background RMS. Notes from visual inspection of the source, if blank then the source appears as a point source with no nearby sources. If notes contains an integer + letter, i.e. 55A, this denotes that the source could be associated with other nearby sources of with the same note. The full catalog contains additional columns. }
	\label{table:Catalog_example}
\end{table*}



\begin{figure*}
  \includegraphics[width=\textwidth]{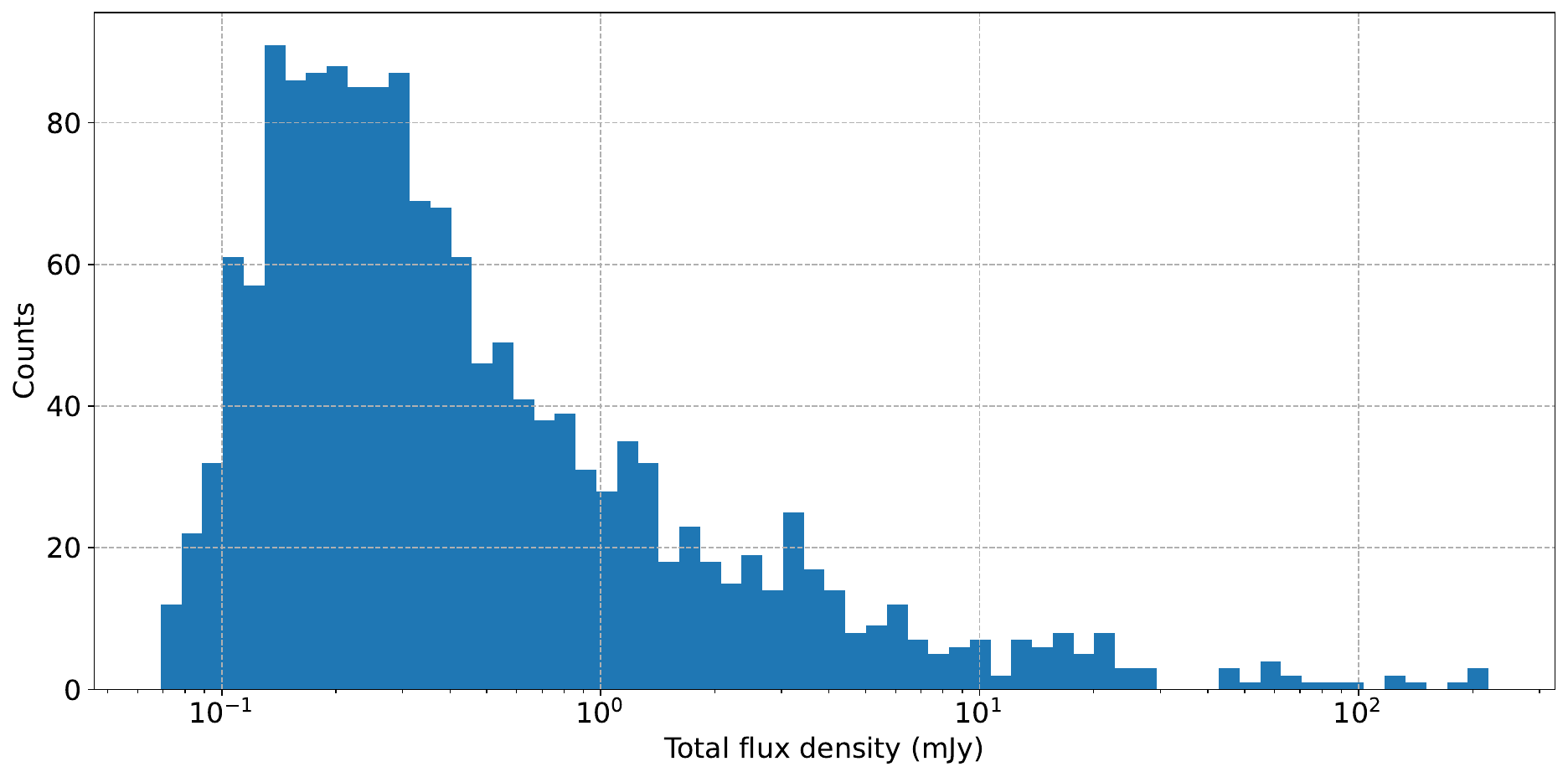}
  \caption{Histogram of flux densities of all sources. Note that there is an effect from the primary beam sensitivity at low flux densities, beginning at 0.6~mJy (estimated from Table~\ref{tab:sensitivity}).}
\end{figure*}

\begin{figure*}
  \includegraphics[width=\textwidth]{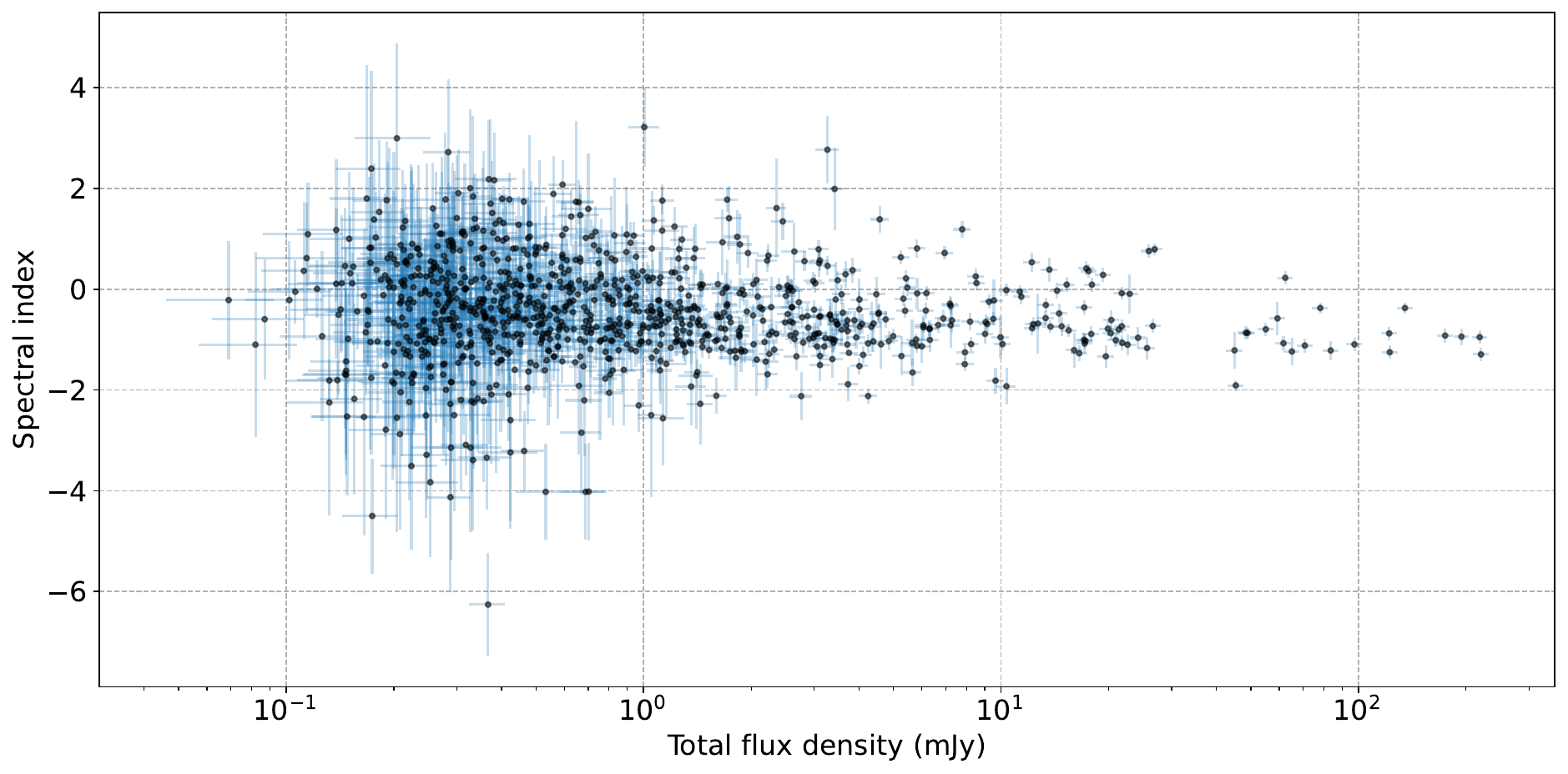}
  \caption{Spectral indices and fluxes of sources which have a spectral index calculated.}
\end{figure*}

\begin{figure*}
  \includegraphics[width=\textwidth]{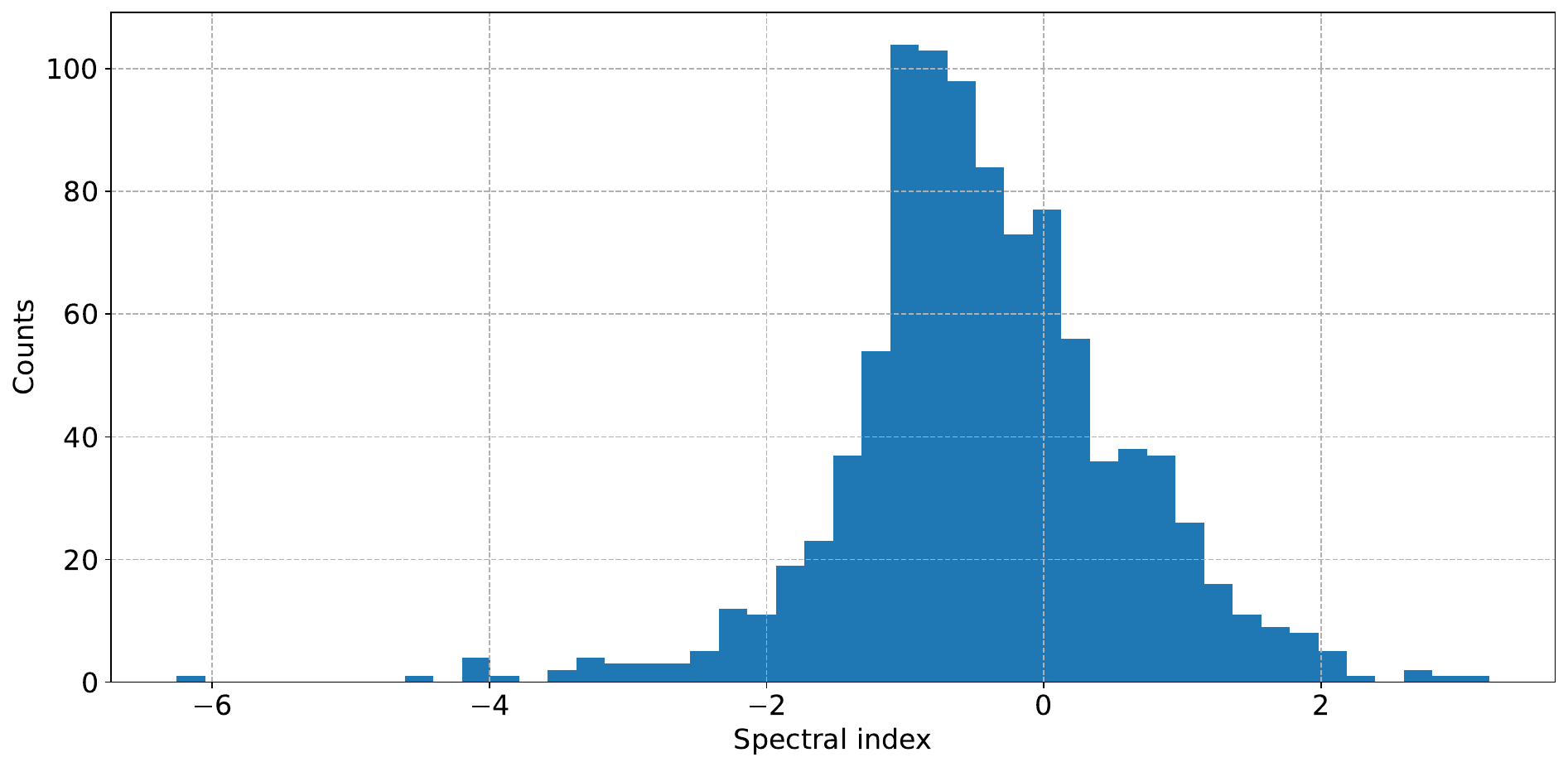}
  \caption{Histogram of spectral indices.}
\end{figure*}

\begin{figure*}
  \includegraphics[width=\textwidth]{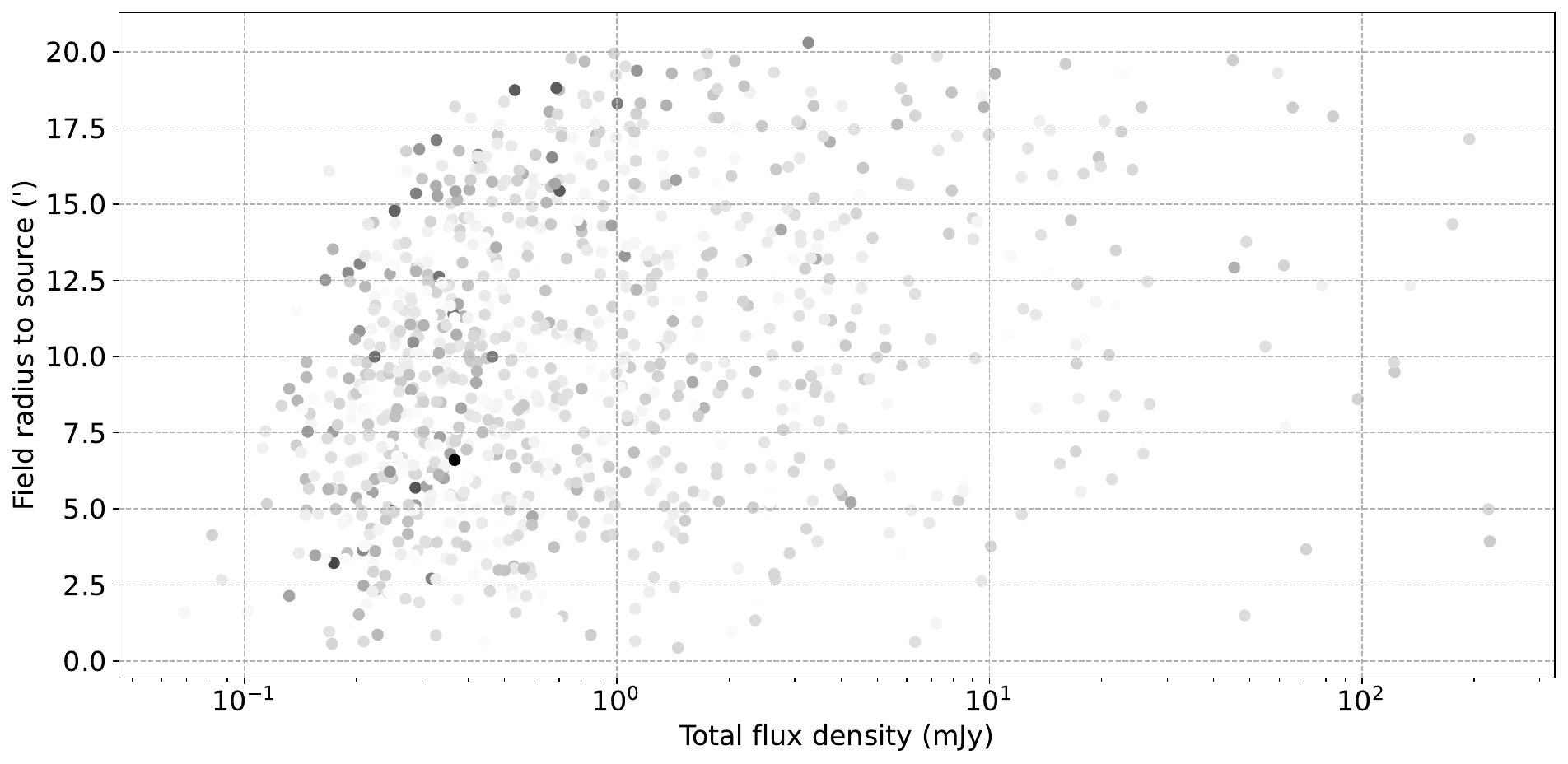}
  \caption{Plot of distance from the centre of the field to source's location and total flux. Points are colored by the absolute value of the spectral index, $| \alpha |$, where darker colors are higher $| \alpha |$ values. We note that a majority of dark points are near the edge of the flux-field distance relation (which is curved due to the effect of the primary beam of decreased sensitivity further from the centre of the field). A similar trend  of more extreme spectral indices with distance from pointing centre was noted in \citet{Smolcic2017}.}
  \label{fig:spedIndIssue}
\end{figure*}

\begin{figure*}
\begin{tabular}{cc}
\subfloat[Artifact amplification source in field 70]{\includegraphics[width=0.45\textwidth]{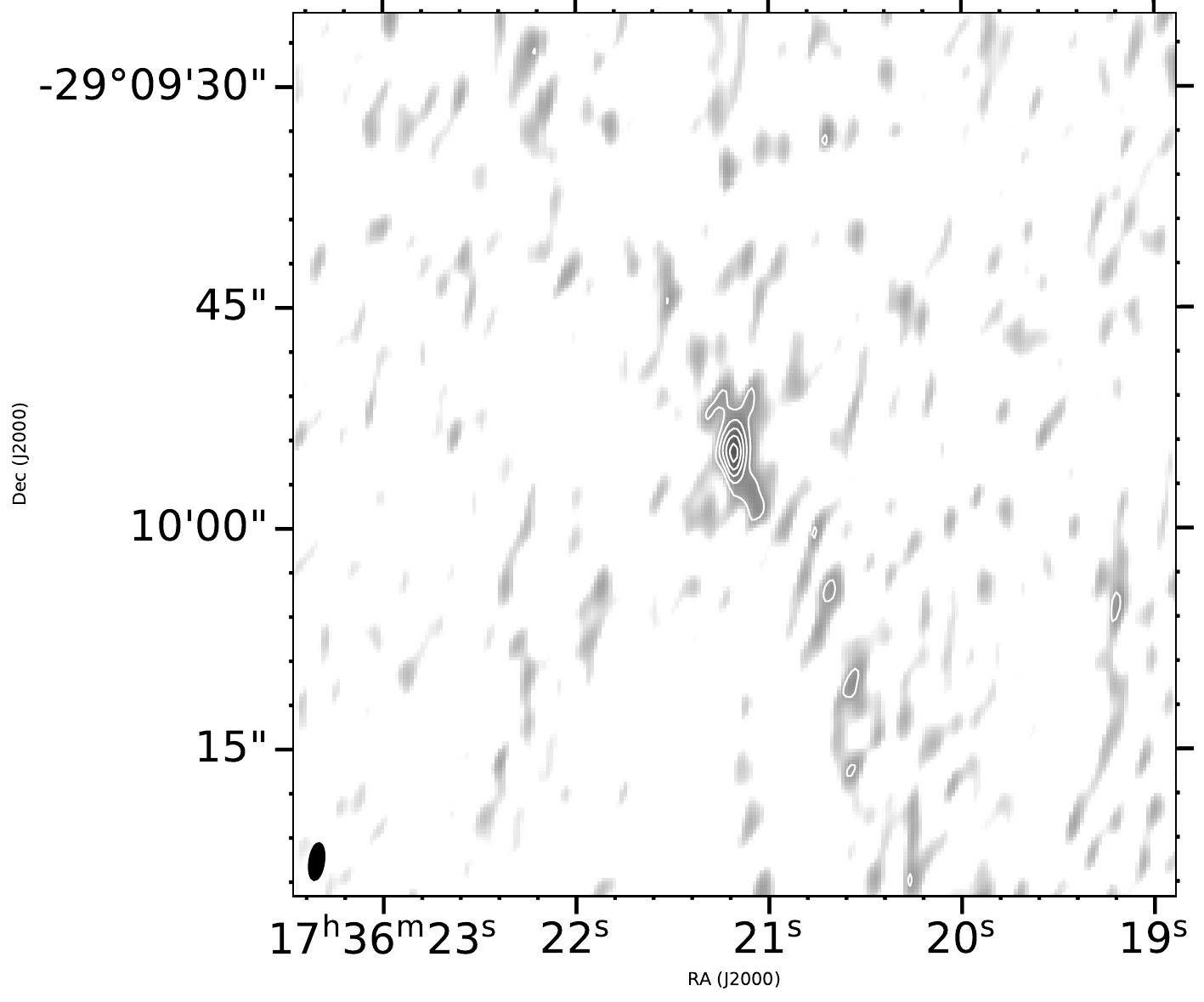}} &
\subfloat[Artifact amplification source in field 103]{\includegraphics[width=0.45\textwidth]{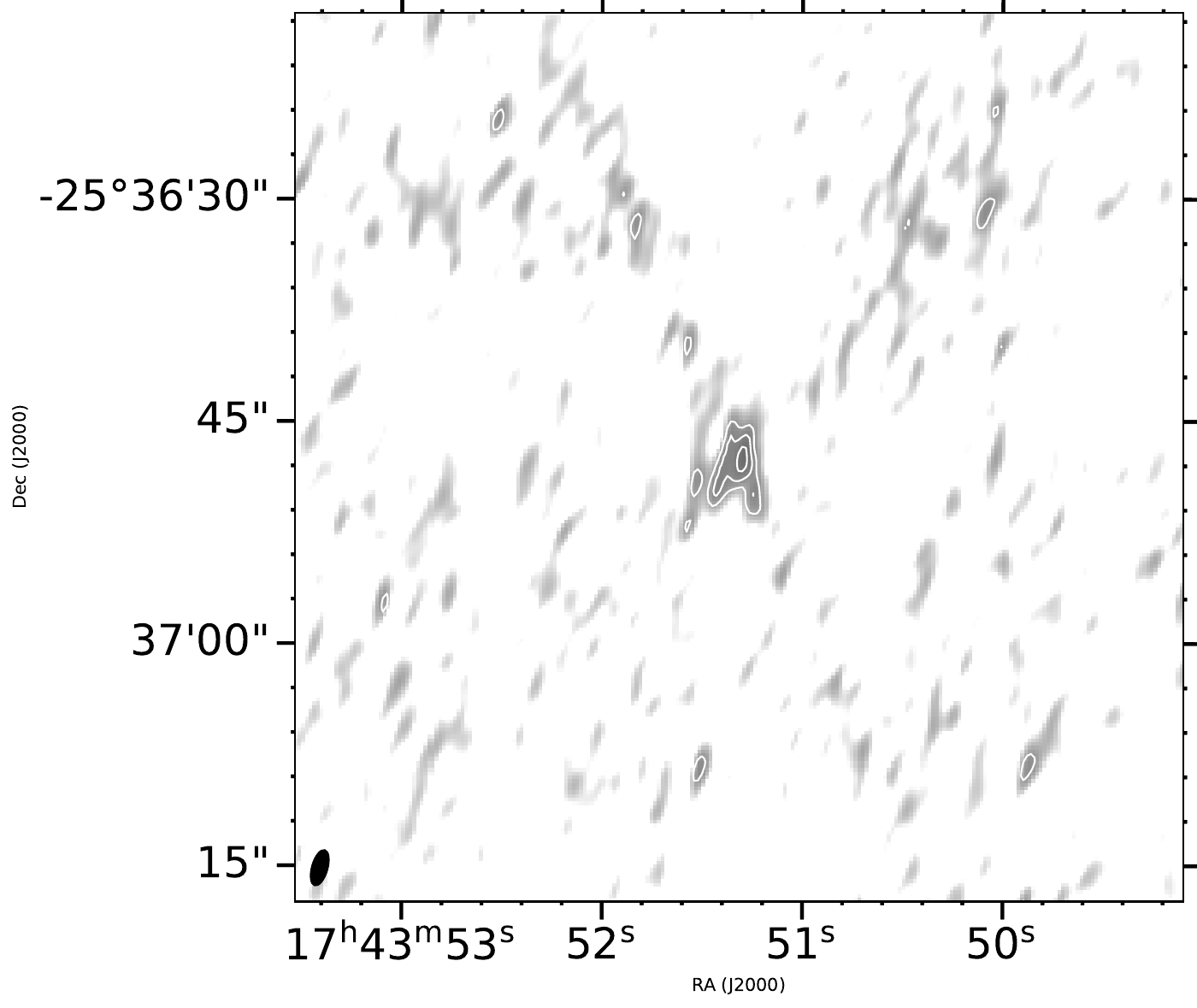}} \\
\end{tabular}
\caption{Two sources that we believe to be randomly amplified manifestations of artifact interference as mentioned in Section~\ref{sec:specIndErr}. Some of the faint artifact lines can be seen intersecting each source (particularly running $\sim$northeast-southwest in direction). Contours begin at 3$\sigma$ (3•RMS), increasing by 1$\sigma$ per level.}
\label{image:artifactAmpSources}
\end{figure*}

\begin{figure*}
\begin{tabular}{c}
\subfloat[VLAGBS field overlap position offset (primary - secondary entry position)]{\includegraphics[width=0.98\textwidth]{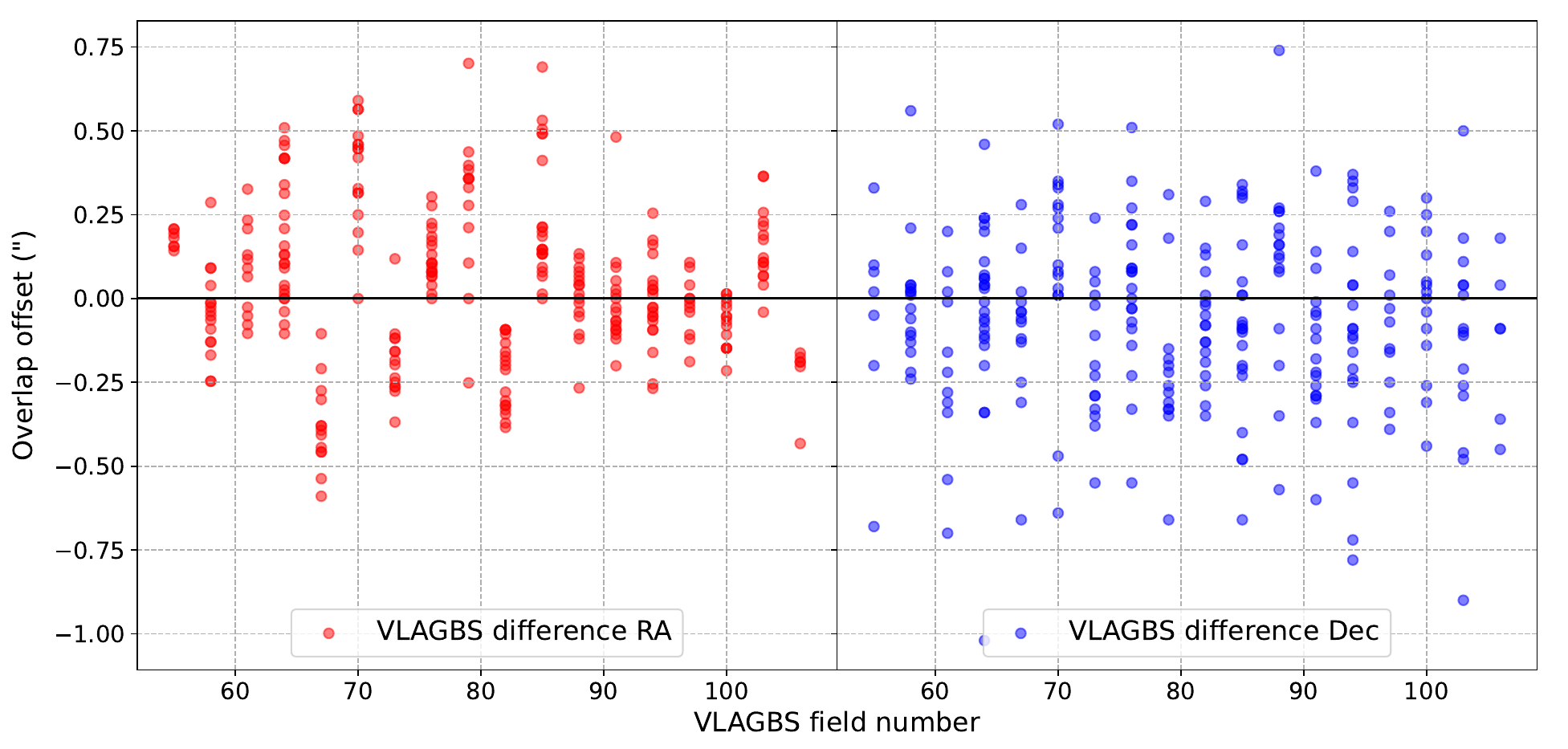}} \\
\subfloat[VLASS position offset (VLAGBS - VLASS position)]{\includegraphics[width=0.98\textwidth]{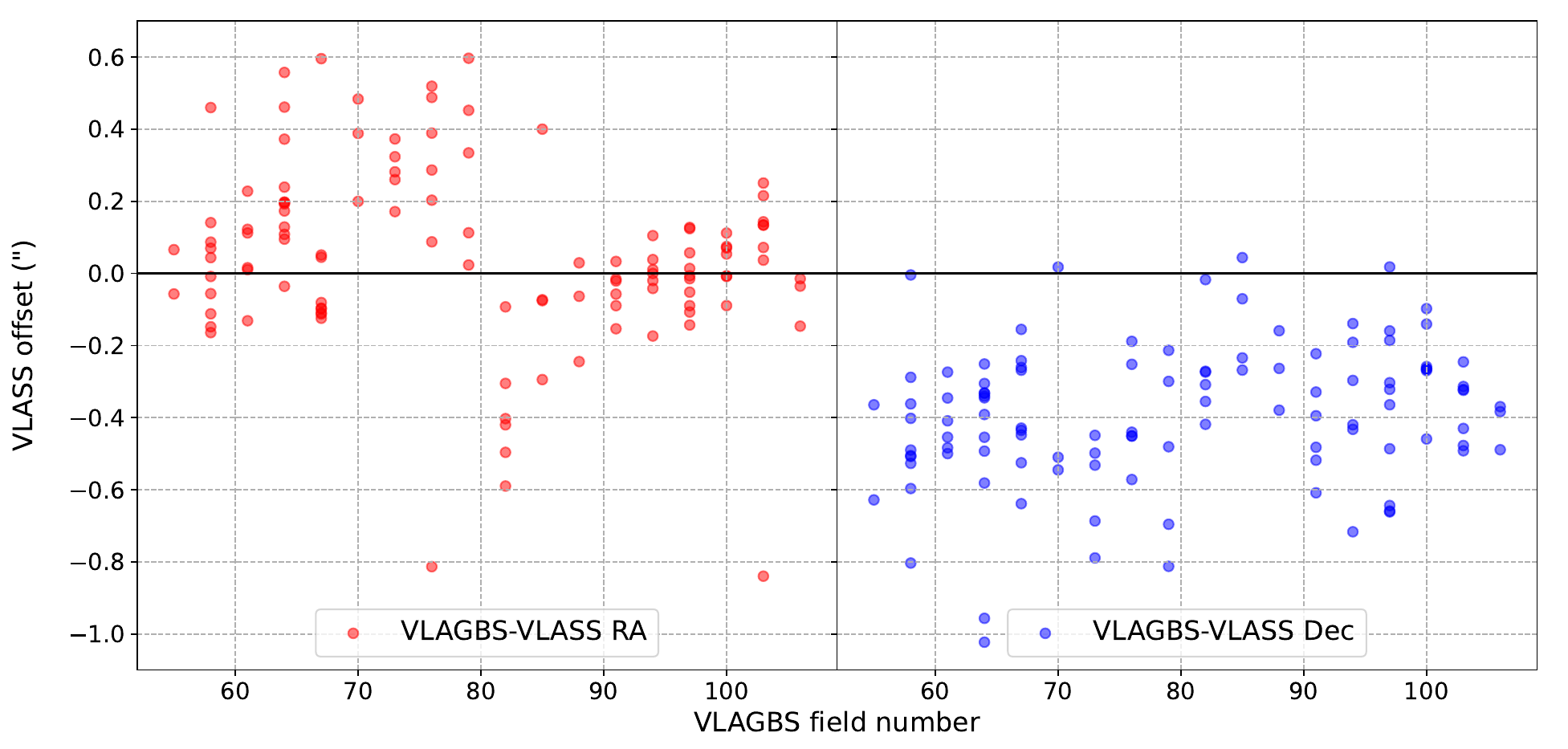}} \\
\end{tabular}
\caption{Comparison of the positions of VLAGBS sources using two methods. Top is the RA and DEC of VLAGBS sources in overlapping fields, and bottom are sources in both VLAGBS and VLASS. All sources used are point sources in VLAGBS (and VLASS for the bottom plot).}
\label{image:positioncompare}
\end{figure*}


\bsp	
\label{lastpage}
\end{document}